\documentclass[12pt,a4paper]{article}
\usepackage{graphicx,color}
\usepackage{amsmath}
\usepackage{amssymb}
\pdfoutput=1
\usepackage{multirow}
\usepackage{url}
\usepackage{subfigure}
\usepackage{authblk}
\usepackage{fullpage}
\usepackage{lscape}
\begin{document}
\vspace*{-0.5in}
\thispagestyle{empty}

\vspace{0.5in}
\begin{center}
\begin{LARGE}
{\bf Status Report:\\
Searching for a Sterile Neutrino at J-PARC MLF (E56, JSNS$^2$)\\}
\vspace{5mm}
\end{LARGE}
\begin{large}
\today\\
\end{large}
\vspace{5mm}
{\large
M.~Harada, S.~Hasegawa, Y.~Kasugai, S.~Meigo,  K.~Sakai, \\
S.~Sakamoto, K.~Suzuya \\
{\it JAEA, Tokai, JAPAN}\\
\vspace{2.8mm}
E.~Iwai, T.~Maruyama\footnote{{Spokesperson: 
(takasumi.maruyama@kek.jp) }}, 
S.~Monjushiro, K.~Nishikawa, M.~Taira\\
{\it KEK, Tsukuba, JAPAN}\\
\vspace{2.8mm}
M.~Niiyama\\
{\it Department of Physics, Kyoto University, JAPAN}\\
\vspace{2.8mm}
S.~Ajimura, T.~Hiraiwa, T.~Nakano, M.~Nomachi, T.~Shima\\
{\it RCNP, Osaka University, JAPAN}\\
\vspace{2.8mm}
T.~J.~C.~Bezerra, E.~Chauveau, H.~Furuta, F.~Suekane\\
{\it Research Center for Neutrino Science, Tohoku University, JAPAN}\\
\vspace{2.8mm}
I.~Stancu\\
{\it University of Alabama, Tuscaloosa, AL 35487, USA}\\
\vspace{2.8mm}
M.~Yeh\\
{\it Brookhaven National Laboratory, Upton, NY 11973-5000, USA}\\
\vspace{2.8mm}
W.~Toki\\
{\it Colorado State University, Fort Collins, Colorado, USA}\\
\vspace{2.8mm}
H.~Ray\\
{\it University of Florida, Gainesville, FL 32611, USA}\\
\vspace{2.8mm}
G.~T.~Garvey, C.~Mauger, W.~C.~Louis, G.~B.~Mills, R.~Van~de~Water\\
{\it Los Alamos National Laboratory, Los Alamos, NM 87545, USA}\\
\vspace{2.8mm}
J.~Spitz\\
{\it University of Michigan, Ann Arbor, MI 48109, USA}
}
\end{center}

\renewcommand{\baselinestretch}{2}
\large
\normalsize

\setlength{\baselineskip}{5mm}
\setlength{\intextsep}{5mm}

\renewcommand{\arraystretch}{0.5}

\newpage

\tableofcontents
\vspace*{0.5in}
\setcounter{figure}{0}
\setcounter{table}{0}
\indent

\section{Introduction}
\indent

The JSNS$^2$ (J-PARC E56) experiment aims to search for sterile neutrinos at the
J-PARC Materials and Life Sciences Experimental Facility (MLF).
After the submission of a proposal~\cite{CITE:PROPOSAL} to the J-PARC PAC, stage-1 approval was granted to the JSNS$^2$ experiment.
The approval followed a series of background measurements which were performed in 2014~\cite{CITE:SR_14NOV}.

Subsequent for stage-1 approval, the JSNS$^2$ collaboration has made continuous efforts to
write a Technical Design Report (TDR).
This TDR will include two major items as discussed in the previous
status report for the 20th J-PARC PAC~\cite{CITE:20PAC}:
\begin{itemize}
\item A realistic detector location
\item Well understood and realistic detector performance using simulation studies, primarily in consideration of fast neutron rejection.
\end{itemize}
Since August we have been in discussions with MLF staff regarding an appropriate detector location.  We are also in the process of
setting up a Monte Carlo (MC) simulation framework
in order to study detector's
performance in realistic conditions. In addition, we have pursued hardware R$\&$D work for the liquid scintillator (LS)
and to improve the dynamic range of the 10'' photomultiplier tubes (PMTs).
The LS R$\&$D works includes Cherenkov studies inside the LS, and a Pulse Shape
Discrimination (PSD) study with a test-beam, performed at Tohoku
University. 
We also estimate the PSD performance of a full-sized detector using a detailed
MC simulation.

In this status report, we describe progress on this work.

\section{Discussion for the detector location with MLF}
\indent

\begin{figure}
\begin{center}
\includegraphics[width=10cm]{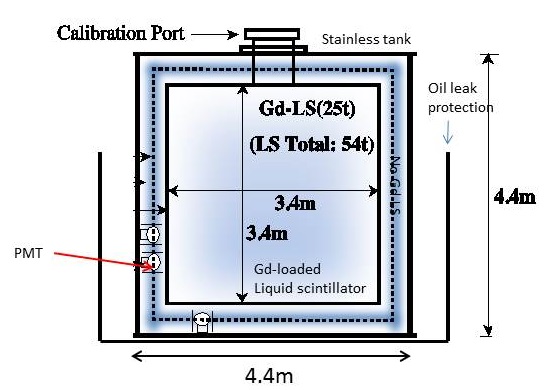}
\caption{A schematic of a JSNS$^2$ detector}
\label{Fig:JSNS2det}
\end{center}
\end{figure}

\begin{figure}
\begin{center}
\includegraphics[width=9cm]{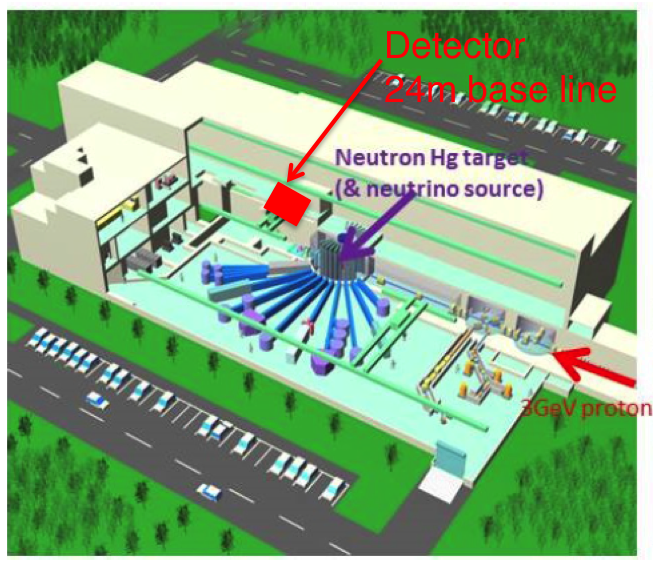}
\caption{Candidate site for JSNS$^2$ detectors}
\label{JSNS2candidate}
\end{center}
\end{figure}

To maximize sterile neutrino sensitivity in the region of interest, it is optimal to put the
JSNS$^2$ detectors (Fig.~\ref{Fig:JSNS2det}) 24~meters from the MLF mercury target, as reported in
Ref.~\cite{CITE:SR_14NOV}.
This position is located in the ``large component handling room'',
the room to handle large equipment of the MLF, 
on the third floor in the MLF building (Fig.~\ref{JSNS2candidate}).
The room is not a common experimental area, but  
is considered a maintenance work area for MLF operations.
Therefore, in order to carry out the JSNS$^2$ experiment, we need to discuss and understand the issues associated with this detector location.
In addition, it is important to clarify the application of safety regulations for the
JSNS$^2$ detectors, which contains more than 100~tons of liquid
scintillator.

In August, 2015, the JSNS$^2$ collaboration requested the KEK IPNS director / the director of J-PARC center
facilitate discussions between MLF and JSNS$^2$, to address these issues in detail.  Since that time we have held several
meetings with directors, MLF, and the local fire department.

In the discussion, there are no show-stoppers that inhibit the JSNS$^2$
experiment. 
The JSNS$^2$ collaboration will try to realize the experiment with further
discussions in detail in the TDR.

\section{R$\&$D for the JSNS$^2$ Detector}

\subsection{Cherenkov study with LAB+0.03g b-PBD}
\indent

As described in the previous status report~\cite{CITE:20PAC}, we verified 
the performance of the liquid scintillator used in the
LSND experiment, mineral oil plus 0.03~g/L b-PBD.

In this status report we investigate the performance of LAB + 0.03g/L b-PBD
liquid scintillator since we plan to use the LAB in the JSNS$^{2}$ experiment.
We use the LSND 0.03~g/L b-PBD as a reference point for the LAB performance
checks. The amount
of secondary light emission materials such as b-PBD or PPO determine
the light yield, and the type of material determines the light
emission time, in addition to the LAB properties.

\subsubsection{Setup of the Measurement}
\indent

\begin{figure}[htbp]
\centering
\subfigure[Setup for Cherenkov + scintillation]{
\includegraphics[width=0.45\textwidth,angle=0]{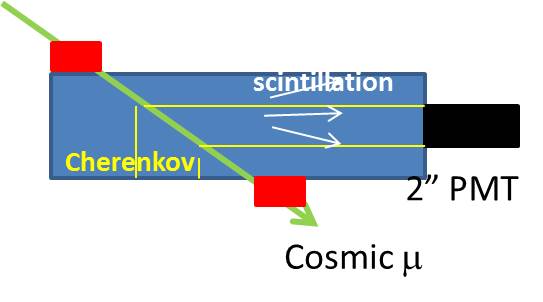}
}
\subfigure[Setup for scintillation light only]{
\includegraphics[width=0.44\textwidth,angle=0]{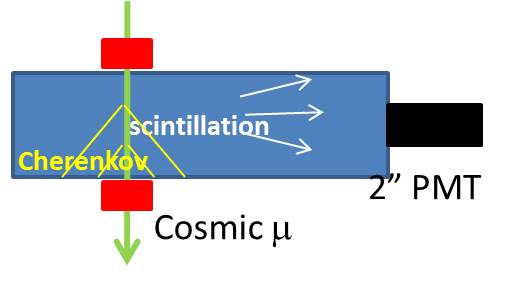}
}
\caption{\setlength{\baselineskip}{4mm}
A schematic diagrams of the test setups. (a) shows the setup for the Cherenkov
plus scintillation detection and (b) shows the setup for the scintillation 
light only detection. The cylindrical prototype with 
a size of 130~mm (diameter) $\times$ 1000~mm (height) is filled with the liquid
scintillator with LAB plus 0.03~g/L b-PBD (blue regions in the schematics).
Scintillation counters, which are shown as red boxes, 
are used to tag cosmic ray muons. Coincidence signals by the scintillation 
counters are used as the reference timing for the Cherenkov and scintillation 
light inside the cylinder.
The effective area for each scintillator
is 5 $\times$ 10~cm$^2$. The average muon path distance from the 2~inch PMT
was kept as 65~cm to make the amount of scintillation light the same in both settings.
} \
\label{FIG:SETUPCS}
\end{figure}

The setup and the data analysis of the measurement is exactly the same as that
in the previous status report~\cite{CITE:20PAC} (Fig.~\ref{FIG:SETUPCS}).
The cylindrical prototype is filled with the liquid scintillator, and a black
sheet is used to avoid reflections of light inside the cylinder.
The motivation of this experiment is to mainly measure the timing difference 
between Cherenkov light and scintillation light. In general, the Cherenkov light 
has a faster emission time than that of the scintillation light. 

\subsubsection{Data Analysis and Results}
\subsubsection*{Hit definition}
\indent

\begin{figure}[htpb]
 \centering
 \includegraphics[width=0.6 \textwidth]{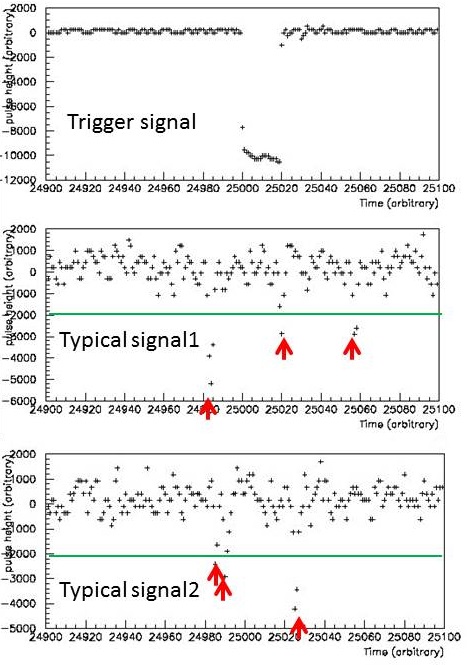}
\caption{\setlength{\baselineskip}{4mm}
Typical timing distributions of the coincidence from the scintillation counters (top) and scintillation light (middle and bottom) measured by the oscilloscope. The horizontal axes corresponds to the time. One count is 2~ns. The vertical axes of the plots is the pulse height counts. 1000~counts corresponds to 1~mV in the bottom plot for this case.
}
 \label{Fig:raw_timing}
\end{figure}

Figure~\ref{Fig:raw_timing} shows a set of typical waveform timing distributions
measured by a digital oscilloscope. We used the oscilloscope data for this
analysis.
The top plot shows typical timing distributions of the coincidence from 
the scintillation counters, and the middle and bottom plots show typical
scintillation light signals, which corresponds to the signals with the setup of
Fig.~\ref{FIG:SETUPCS}(b). For the explanation of the analysis, only the
typical signals in Fig.~\ref{FIG:SETUPCS}(b) are described here.
Note that the coincidence signal from the scintillation counters was 
delayed due to electronics. Also, this relative timing was smeared by the 
PMT's jitter ($\sim$1 ns), electronics (within 1 ns), and hit position of the
cosmic rays ($<<$ ns). The fastest scintillation light or Cherenkov light should
be located around 24980~time counts. This was demonstrated by
using a Water Cherenkov setup~\cite{CITE:20PAC}.

As shown in Fig.~\ref{Fig:raw_timing}, there is a threshold to determine the
timing of the ``hit'', which is 2000~counts ($\sim$ 1/3~p.e.) in the data.
In the figure, thresholds correspond to the green lines, and the red arrows
show the recognized hits above the thresholds. These typical signals have multiple hits.

\subsubsection*{\setlength{\baselineskip}{4mm}
Results: Comparison between scintillation only and Cherenkov+scintillation light}
\indent

\begin{figure}[htpb]
 \centering
 \includegraphics[width=0.6 \textwidth]{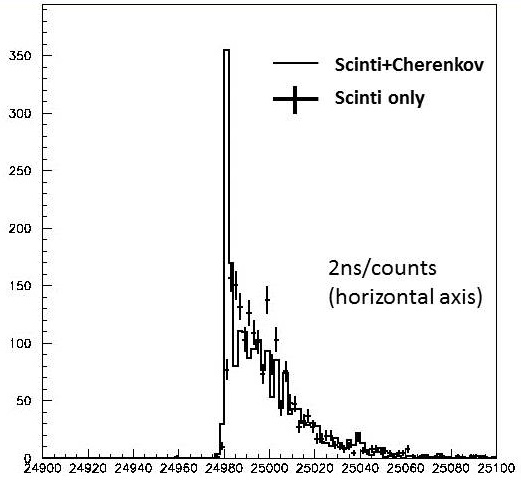}
\caption{\setlength{\baselineskip}{4mm}
Timing comparison between scintillation light only (cross) and Cherenkov + scintillation light (histogram). Cherenkov light is concentrated in the fast timing region.}
 \label{Fig:timing}
\end{figure}

Figure~\ref{Fig:timing} shows the relative timing of hits
with respect to the coincidence defined by the thresholds, and also shows 
the timing comparison between scintillation light only (cross) and Cherenkov +
scintillation light (histogram).
As clearly seen, Cherenkov light is concentrated in
the fast timing peak. If we select the fastest timing in the first 2~ns bin, the
$\sim$ 95$\%$ of scintillation light is rejected from Cherenkov light
while only $\sim$25$\%$ of scintillation light is contaminated into the
Cherenkov light.

An estimated light yield is $\sim$1600~photons/MeV. Compared to the
$\sim$9000~photons/MeV measured by Daya Bay experiment~\cite{CITE:DayaBayLS},
a dilution factor, which is defined as the ratio between the light yield of
the nominal scintillator and that of the scintillator with small amount of
the secondary light emission materials, is 5.3.

\subsubsection{Simulation of the Measurement}
\indent

\begin{figure}[hbtp]
	\begin{center}
		\includegraphics[scale=0.6]{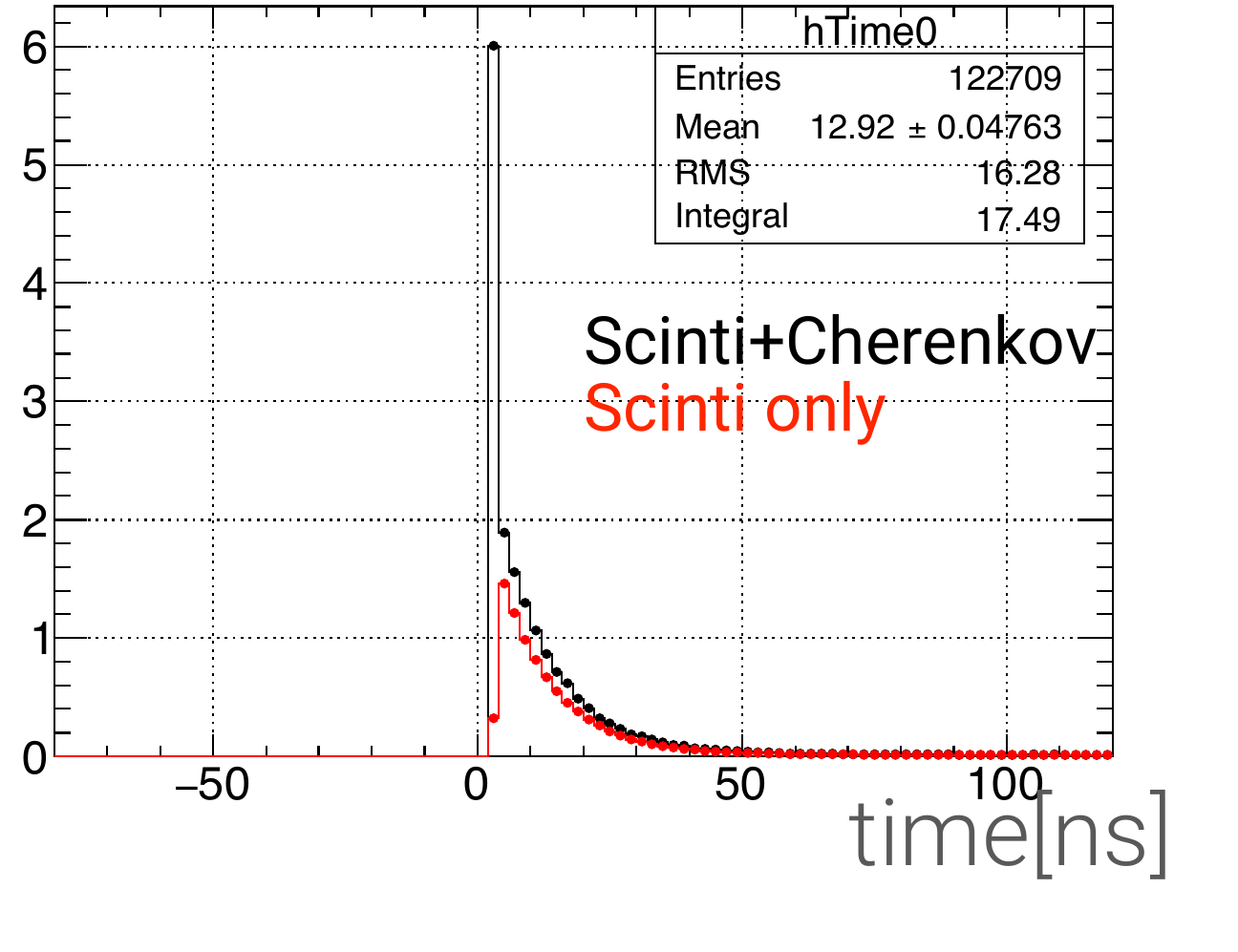}
		\caption{\setlength{\baselineskip}{4mm} Timing distribution of single photoelectrons for scintillation+Cherenkov light (black) and scintillation light only (red), obtained with a MC simulation.}
		\label{fig_timing}
	\end{center}
\end{figure}

We also evaluated the results using a simulation to understand the optical properties of the medium.
The software will be described later in Section~\ref{SEC:SW}.
Figure~\ref{fig_timing} shows the timing distribution of single photoelectrons
obtained with the Monte Carlo simulation before parameter tuning.
We will continue to study the measurement results to tune the optical parameters precisely.

\subsubsection{\setlength{\baselineskip}{4mm}
Results; Comparison between Mineral Oil and LAB}
\indent

\begin{figure}[htpb]
 \centering
 \includegraphics[width=0.8 \textwidth]{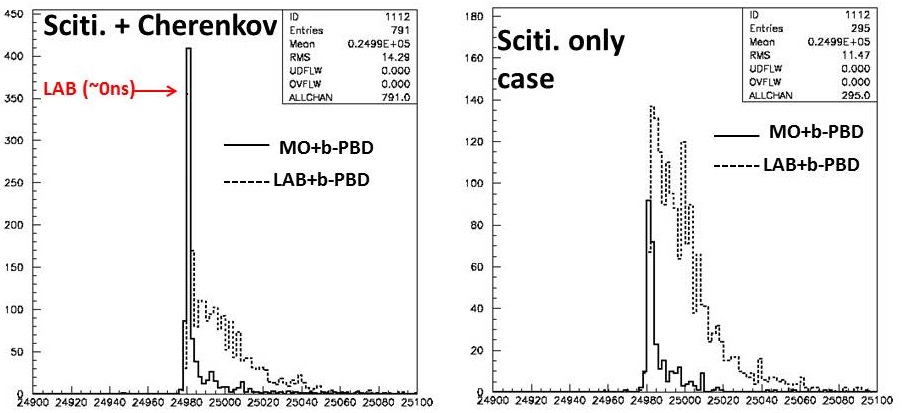}
\caption{\setlength{\baselineskip}{4mm}
Timing comparison between LAB and mineral oil. Cherenkov + scintillator (Left) and Scintillator only (Right).}
 \label{Fig:MOvsLAB}
\end{figure}

Finally, we compare the results of this study between mineral oil and LAB (Fig.~\ref{Fig:MOvsLAB}).
The following observations have been made: (A) Emission time of scintillation light
with LAB+0.03~g/L b-PBD is
wider than that in the mineral oil+0.03~g/L b-PBD. (B) Light yield in the 
LAB+0.03~g/L b-PBD is higher than that in the mineral oil+0.03~g/L b-PBD case
by a factor of six. (C) The yield of the Cherenkov light is the same in the two
liquid scintillators within 10$\%$. 

\subsection{Study of PSD capability with JSNS$^{2}$ detector}
\label{SEC:PSD}
\indent

As presented in the previous status report for the PAC~\cite{CITE:20PAC}, there are indications that the
Daya Bay type of Gd-loaded liquid scintillator (DBLS) produces a strong PSD capability in the several MeV
range. The PSD capability is crucial in rejecting cosmic induced fast neutrons by a factor of 100, and
also important for keeping the neutrino detection efficiency as high as possible.
However, the measurements were done with a small vial detector and a less than 10~MeV source ($^{252}$Cf
neutron source). Therefore, it is important to show the feasibility in the actual energy range and full
JSNS$^2$ detector size. In order to demonstrate the PSD capability of the DBLS, a measurement using a 70~MeV
neutron beam created by a cyclotron in the Cyclotron and Radioisotope Center (CYRIC) at Tohoku University
was performed in October 2015. The current status of the MC study and first look plots of the neutron
beam measurement are shown in this section.
Another effort to evaluate the PSD capability with the real size detector was made using a realistic MC
simulation. The PSD parameters were extracted from the vial test, and they were extrapolated to a real
size detector. The results of this study are shown in the following sections.

\subsubsection{\setlength{\baselineskip}{4mm}
  MC study of PSD capability of the DBLS with a full size JSNS$^2$ detector}
\indent

The PSD capability of the DBLS with a full size detector was calculated with a MC simulation based
on Geant4~\cite{geant4}. Each PSD related parameter (scintillation light yield, the light emission
time, Birks constant, etc.) of the DBLS was tuned by using measured data
with a 100~mL cylindrical vial with an attached 2~inch PMT.

\begin{figure}[htbp]
\begin{center}
\includegraphics[width=14cm]{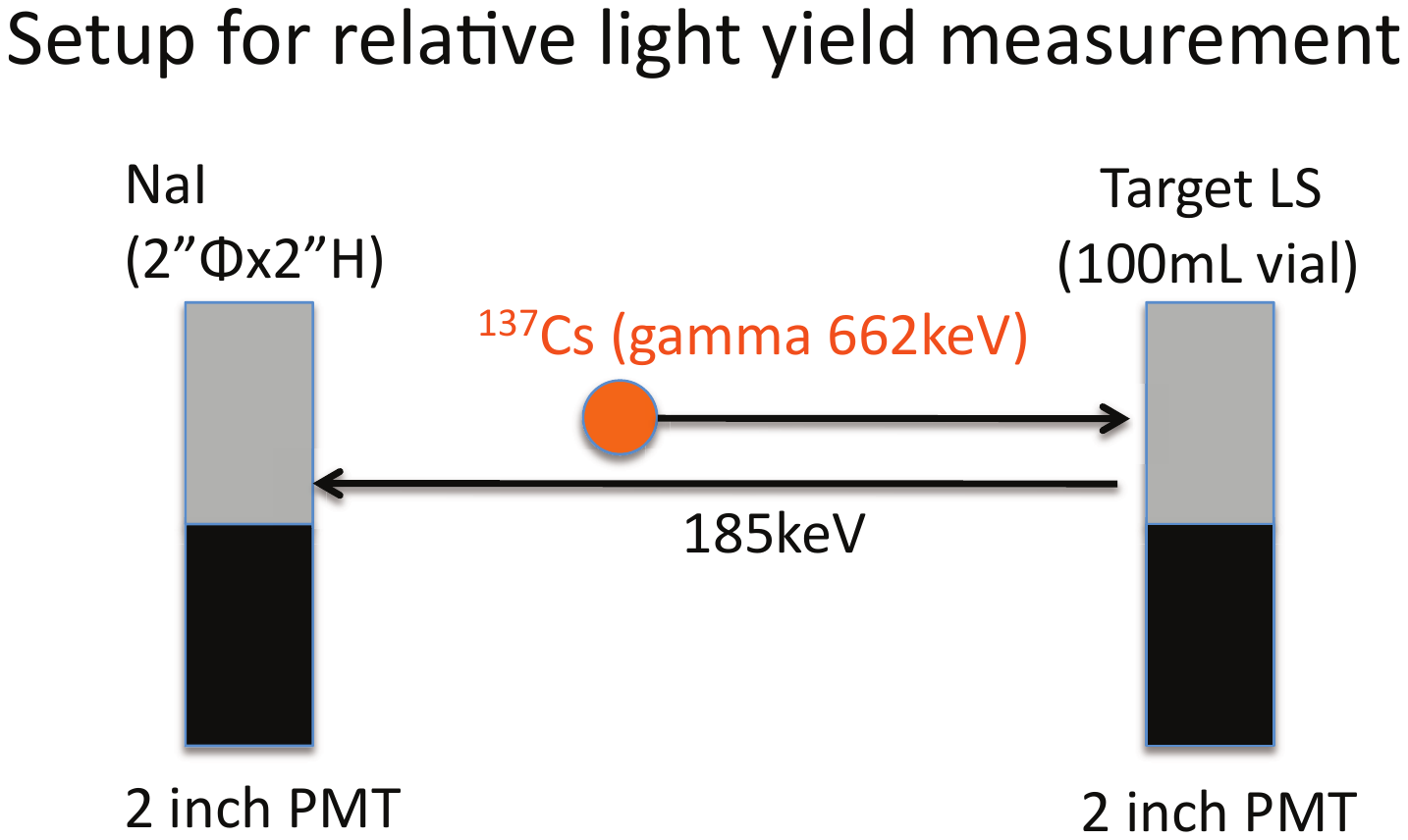}
\caption{\setlength{\baselineskip}{4mm}
Setup for measurement of light yield of the DBLS.}
\label{FIG:CsSetup}
\end{center}
\end{figure}

\begin{figure}[htbp]
\begin{center}
\includegraphics[width=10cm]{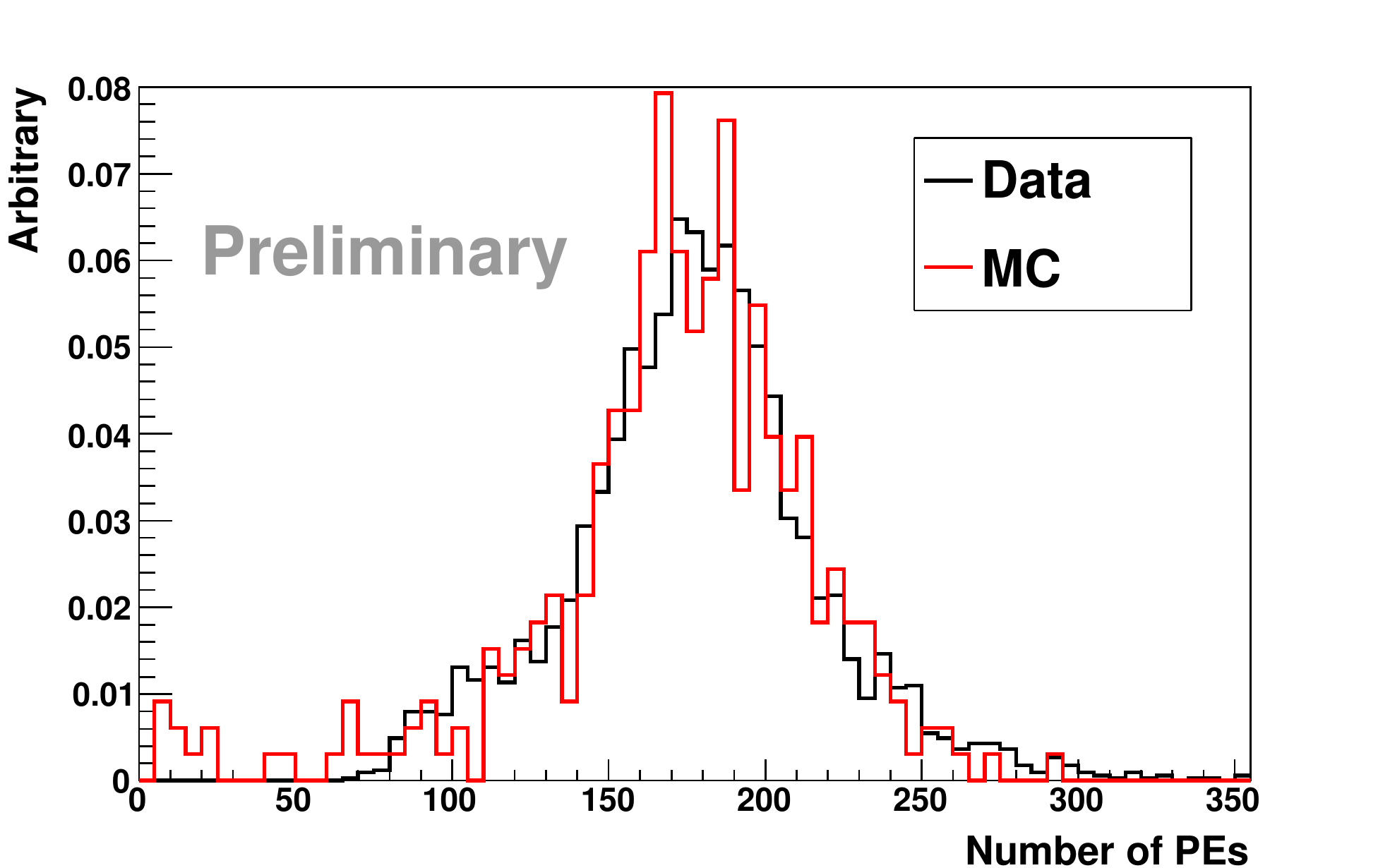}
\caption{\setlength{\baselineskip}{4mm}
  Peaks of the detected number of photoelectrons of the backscattering gamma events from the
  $^{137}$Cs source in the DBLS. The black and red lines show data and MC, respectively.}
\label{FIG:Cs}
\end{center}
\end{figure}

Figure~\ref{FIG:CsSetup} shows the setup for the measurement of light yield of the DBLS.
To detect backscattering events which gamma rays are emitted from a $^{137}$Cs source
and detected by the NaI detector via the Compton scatter in the DBLS detector,
the $^{137}$Cs source is set between the DBLS
and the NaI detectors as shown in Fig.~\ref{FIG:CsSetup}.  This setup allows us to detect
only Compton edge events with scattered angle of 180 degrees. The coincidence signal
from the DBLS and the NaI detectors is used as the trigger. The data are taken with a flash ADC
module (CAEN V1730, 500~MS/s, 14~bits).
The black line in Fig.~\ref{FIG:Cs} shows the number of photoelectrons of the backscattering events.
Comparing the peak of liquid scintillator of KamLAND experiment, with known light yield
(8300~photons/MeV~\cite{CITE:KamLAND}), the relative light yield of the DBLS can be measured.
Thus, the light yield is estimated as $\sim$10000~photons/MeV, compared to the $\sim$9000~photons/MeV
measured by Daya Bay experiment~\cite{CITE:DayaBayLS}.  
The red line in Fig.~\ref{FIG:Cs} shows the MC distribution after implementing the measured light yield.

\begin{figure}[htbp]
\begin{center}
\includegraphics[width=16cm]{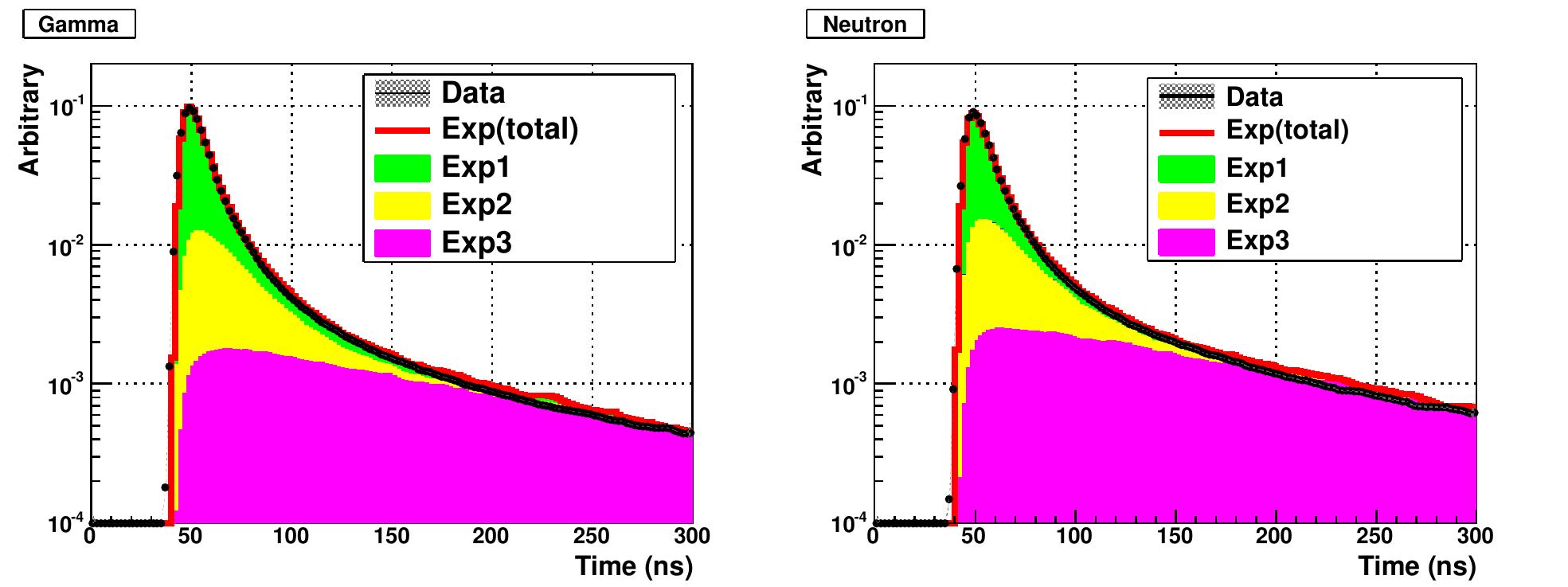}
\caption{\setlength{\baselineskip}{4mm}
  Timing distributions of the events from $^{241}$Am$^{9}$Be source in the DBLS.
  The black and red lines show data and MC, respectively. Left: gamma, Right: neutrons.}
\label{FIG:WFAmBe}
\end{center}
\end{figure}

The black lines in Fig.~\ref{FIG:WFAmBe} show the measured mean waveforms of gammas (left plot)
and neutrons (right plot) from the $^{241}$Am$^{9}$Be source with the vial. The time constants
of the scintillation light emission in the MC simulation were calculated by fitting the mean
waveform data by the expectation, summing 3 exponentials in consideration of the mean waveform
of one photoelectron and TTS of the 2~inch PMT. Table~\ref{TAB:TAmBe} shows each parameter
of the fit result, and filled areas in Fig.~\ref{FIG:WFAmBe} show the expected waveform
of each time constant of the fit result.     

\begin{table}[htb]
\caption{Time constants of the scintillation light emission of the DBLS.}
\begin{center}
\begin{tabular}{|c|c|c|c|} \hline
Component&time constant&ratio (gamma)&ratio (neutron)\\\hline
Exp1&3.87ns&0.673&0.580\\\hline
Exp2&17.6ns&0.187&0.219\\\hline
Exp3&148.3ns&0.140&0.200\\\hline

\end{tabular}
\end{center}
\label{TAB:TAmBe}
\end{table}

\begin{figure}[htbp]
\begin{center}
\includegraphics[width=16cm]{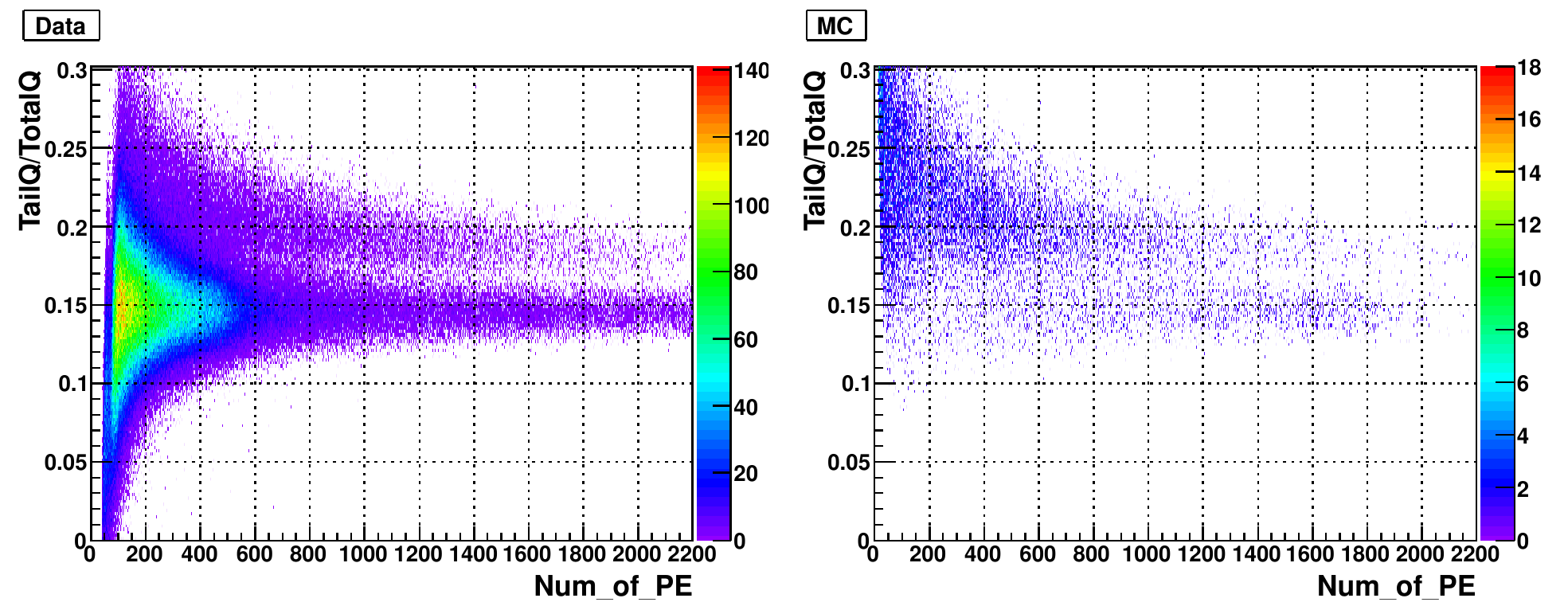}
\caption{\setlength{\baselineskip}{4mm}
  Correlations between the TailQ/TotalQ variable, the definition of which is the same as previous PAC
  report, and the detected number of photoelectrons for the DBLS. The left plot shows data with
  the $^{241}$Am$^{9}$Be source, including backgrounds such as environmental gammas, and the right plot
  shows the MC with only the $^{241}$Am$^{9}$Be source.}
\label{FIG:2DPSDAmBe}
\end{center}
\end{figure}

\begin{figure}[htbp]
\begin{center}
\includegraphics[width=10cm]{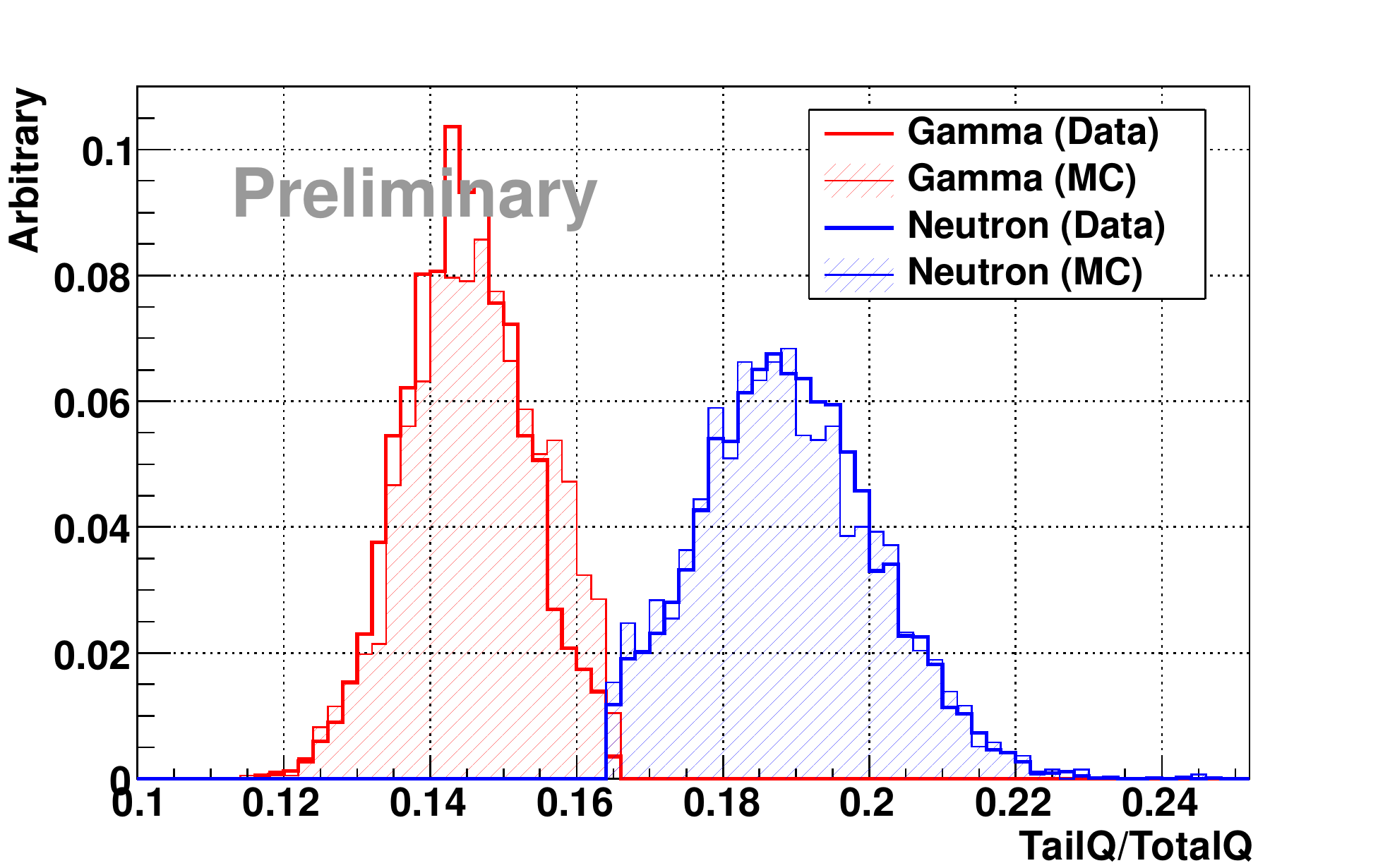}
\caption{\setlength{\baselineskip}{4mm}
The TailQ/TotalQ distributions of the DBLS. The red and blue show gammas and neutrons, respectively. Lines and filled areas show data and the MC, respectively.}
\label{FIG:1DPSDAmBe}
\end{center}
\end{figure}

Figure~\ref{FIG:2DPSDAmBe} shows the TailQ/TotalQ variable, whose definition is the same as the
previous PAC report~\cite{CITE:20PAC}, as a function of the detected number of photoelectrons for the DBLS. The left plot shows data with the $^{241}$Am$^{9}$Be source, including background
such as environmental gammas in particular, and the right plot shows the MC with only
the $^{241}$Am$^{9}$Be source. Although there is a discrepancy due to the environmental
gammas in the range below $\sim$ 600 photoelectrons, the MC reproduces the data well in the
high photoelectron range. Figure~\ref{FIG:1DPSDAmBe} shows the TailQ/TotalQ distributions
of events between 1000 and 2000~photoelectrons. Both the gamma and neutron data are well reproduced by the MC.

\begin{figure}[htbp]
\begin{center}
\includegraphics[width=16cm]{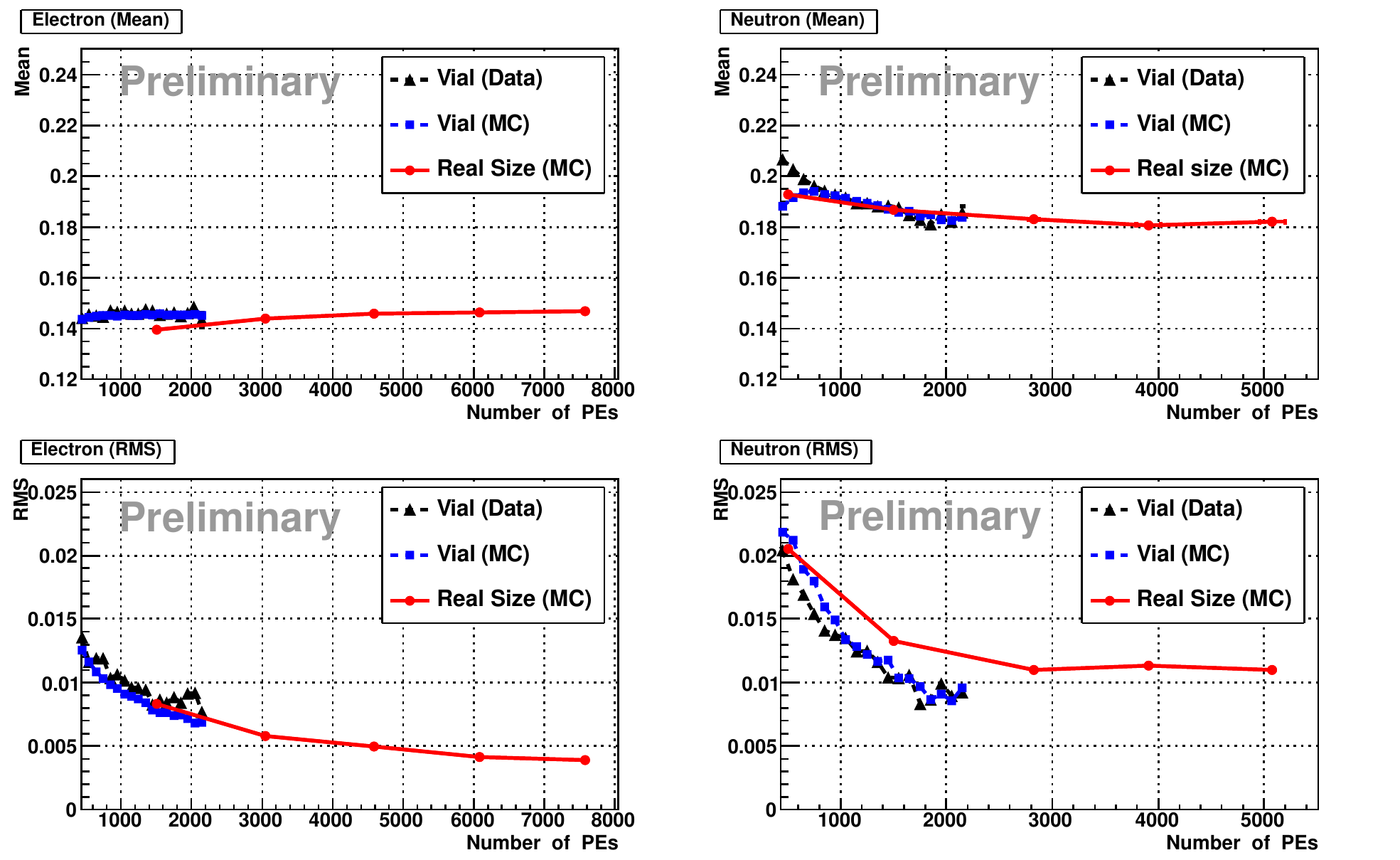}
\caption{\setlength{\baselineskip}{4mm}
  Correlations between peak means of the TailQ/TotalQ distributions and the detected number of photoelectrons
  due to electrons (left) and neutron-induced proton recoils (right). The lower plots show the peak RMS.
  The blue and black markers show the vial measurement data and MC with the $^{241}$Am$^{9}$Be source,
  respectively, and the red lines show MC results with the JSNS$^{2}$ detector.}
\label{FIG:PSDJSNS}
\end{center}
\end{figure}

Using the MC simulation with the tuned parameters mentioned above, the PSD capability of the DBLS with
the realistic JSNS$^{2}$ detector was evaluated. To check the PSD capability in the several tens of MeV
range of the prompt signals, electrons and neutrons with various energies between 10 and 50~MeV were
generated at the JSNS$^{2}$ detector center. The upper figures in Fig.~\ref{FIG:PSDJSNS} show the
means of the TailQ/TotalQ distributions as a function of the detected number of photoelectrons for the
electrons (left) and neutron-induced proton recoils (right). The lower plots show the RMS. The blue and
black markers show the data and MC of the vial measurement with $^{241}$Am$^{9}$Be source, respectively,
and the red lines show MC results with the JSNS$^{2}$ detector.
The vial measurement data are well reproduced by the MC in all plots, and extrapolation curves of the MC
with the JSNS$^{2}$ detector (red lines) from the lower photoelectron range of the vial measurement looks
reasonable, except for the case of RMS of protons recoiled from neutrons (bottom right plot).
The MC with the JSNS$^{2}$ detector is working well. The reason of the discrepancy of the RMS curve
of the recoiled protons is likely due to a geometrical effect; an investigation into this
discrepancy is on-going.

\begin{figure}[htbp]
\begin{center}
\includegraphics[width=16cm]{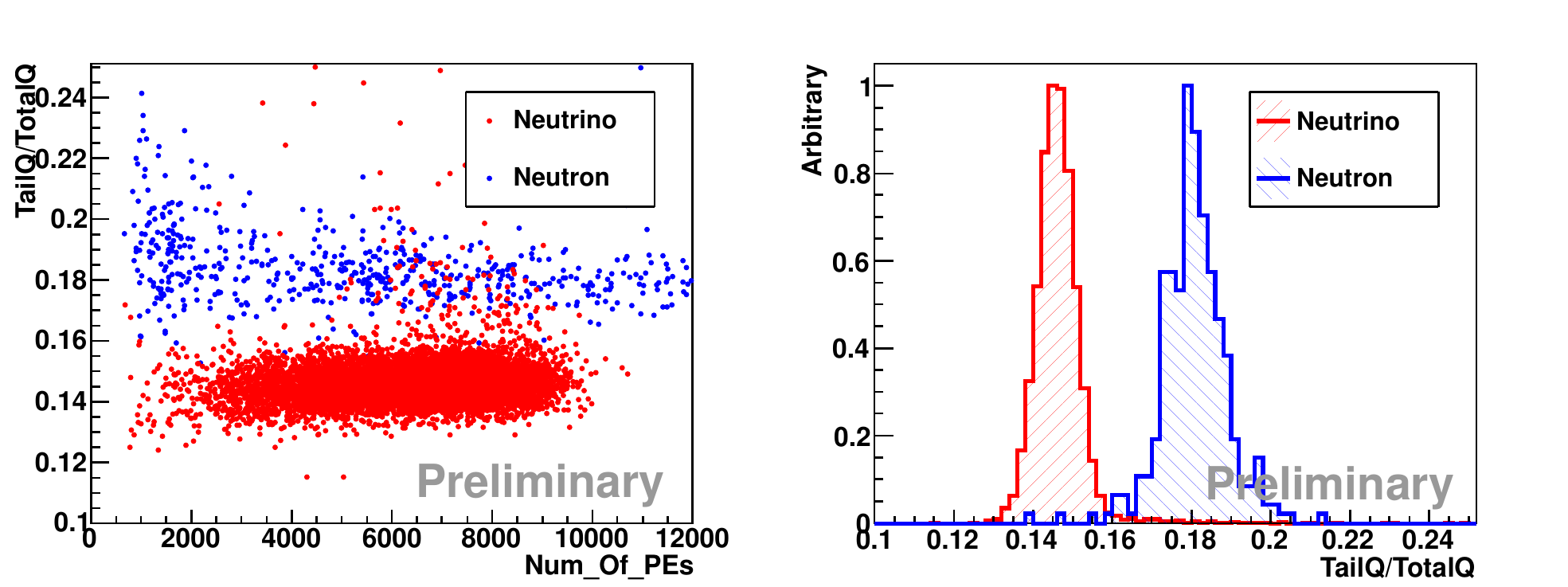}
\caption{\setlength{\baselineskip}{4mm}
  Correlations between peak means of the TailQ/TotalQ distributions and the detected number of photoelectrons
  due to the 100\% oscillated neutrinos and protons recoiled by the cosmic induced fast neurons (left plot).
  The right plot shows the TailQ/TotalQ distributions after applying the neutrino selection criteria. The
  red and blue show the neutrinos and the recoiled protons, respectively.}
\label{FIG:PSDJSNSNu}
\end{center}
\end{figure}

The left plot in Fig.~\ref{FIG:PSDJSNSNu} shows MC results of the correlations between means of the
TailQ/TotalQ distributions and the detected number of photoelectrons due to 100\% oscillated neutrino
signals (red) and protons from cosmic induced fast neutron interactions (blue) generated using the
same method as the proposal for JSNS$^{2}$~\cite{CITE:PROPOSAL}. The right plot shows
the TailQ/TotalQ distributions of the neutrinos and protons after applying a simplified
neutrino selection criteria (20 $\le$ E$_{prompt}$ $\le$ 60 MeV, 5 $\le$ E$_{delayed}$ $\le$ 12 MeV,
0 $\le$ $\Delta$t $\le$ 100 $\mu$sec, multiplicity of thermal neutron capture = 1). The red and blue
colors in Fig.~\ref{FIG:PSDJSNSNu} show the neutrinos and recoiled protons, respectively.
The plot shows good separation between the neutrinos and recoiled protons, due to neutron-proton
elastic scattering. However, there is a possibility to include inelastic scattering or spallation
with gammas between the neutrons and $^{12}$C in the scintillator. If there are such events with gammas,
the neutron rejection power becomes worse compared to the MC result which includes only recoiled protons.
Thus, it is important to check the PSD capability in the events with gammas caused by neutrons
of several tens of MeV.   

\subsubsection{PSD capability with test-beam}
\indent

\begin{figure}[htbp]
\begin{center}
\includegraphics[width=14cm]{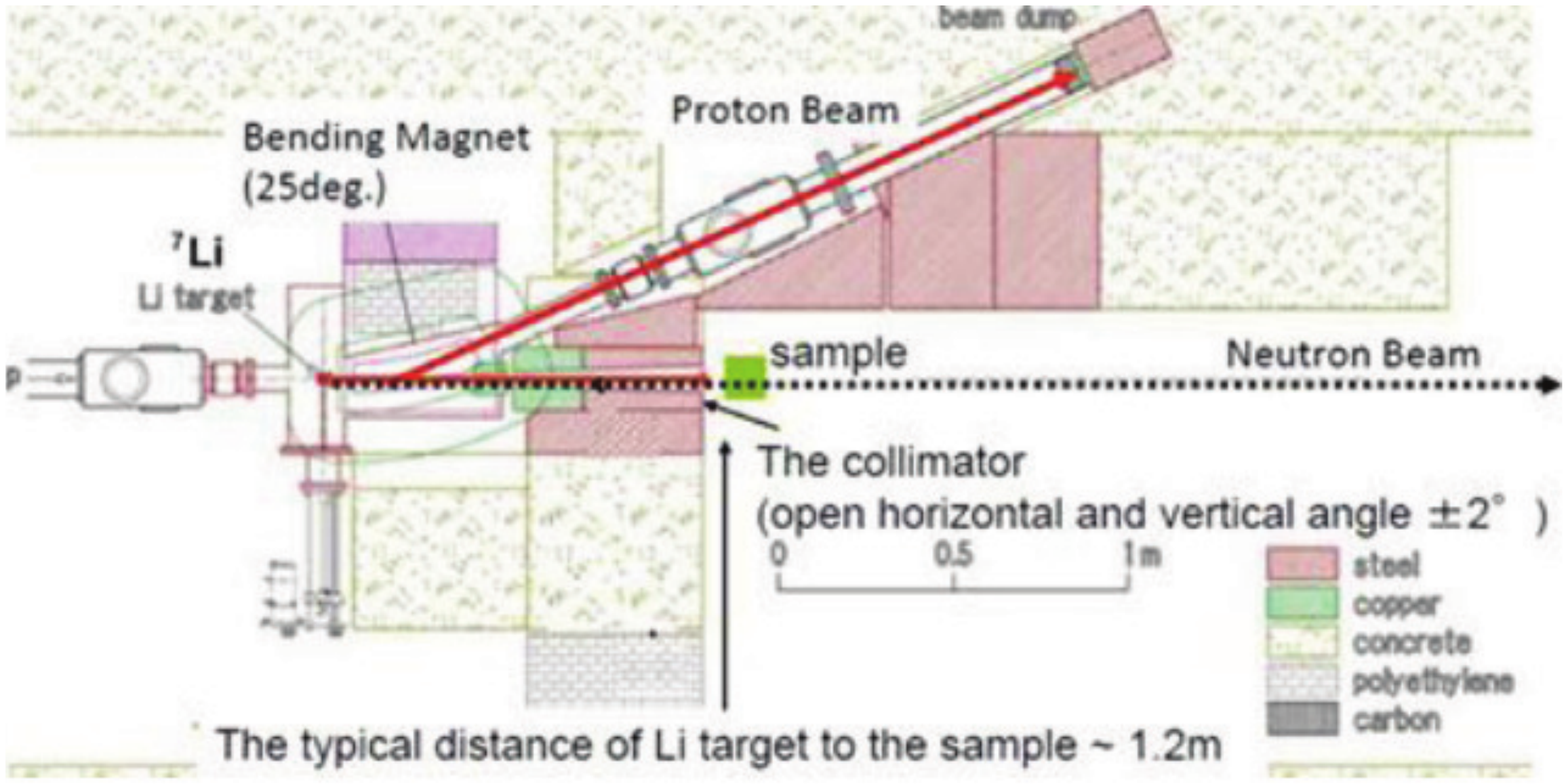}
\caption{\setlength{\baselineskip}{4mm}
Schematic layout of the fast neutron beam facility.}
\label{FIG:CYRIC32}
\end{center}
\end{figure}

To check the PSD capability in the prompt energy range and the inelastic or spallation events between
the cosmic induced fast neutron and $^{12}$C in the scintillator, a measurement with a neutron beam of
70 MeV was carried out in CYRIC in Tohoku University. 
CYRIC has a high-intensity fast neutron facility~\cite{CITE:CYRIC32}. A proton beam provided by the
AVF cyclotron hits a Li target, producing the quasi-monoenergetic neutron beam from 14 to 80 MeV
via the $^{7}$Li(p,n)$^{7}$Be reaction. Figure~\ref{FIG:CYRIC32} shows a schematic layout of the
fast neutron beam facility. The DBLS with the 100~mL vial with an attached 2~inch PMT is set on the
beam axis at the facility. Two cylindrical NaI counters (2" (diameter)$\times$ 2" (height) and 3" (diameter) $\times$ 3" (height)) were also set
on both sides of the DBLS for checking the events following gammas due to the inelastic or spallation
reaction of $^{12}$C. 

\begin{figure}[htbp]
\begin{center}
\includegraphics[width=10cm]{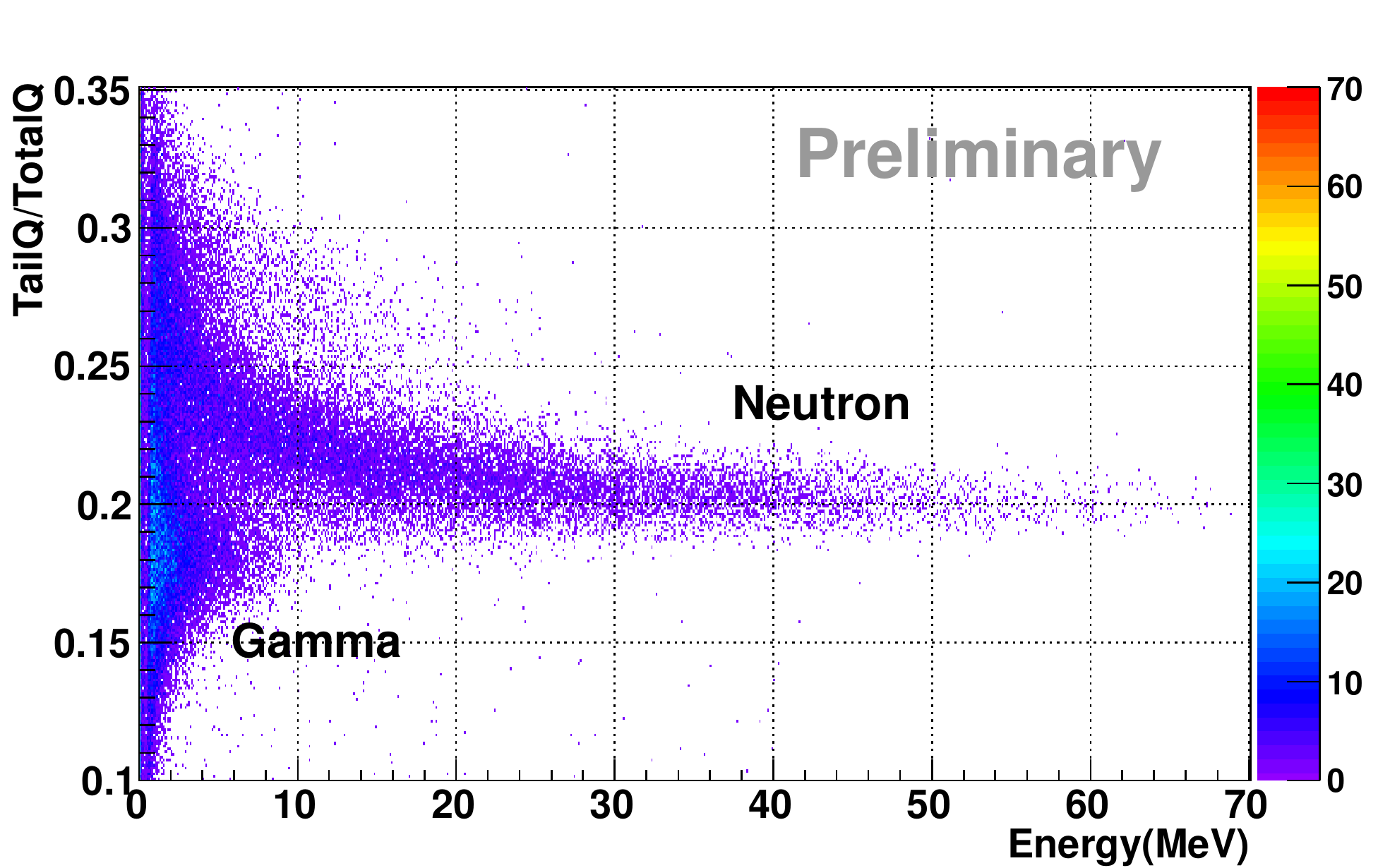}
\caption{\setlength{\baselineskip}{4mm}
The correlation between the TailQ/TotalQ variable and visible energy of the DBLS.}
\label{FIG:BeamPSD}
\end{center}
\end{figure}

Figure~\ref{FIG:BeamPSD} shows a first look plot of the correlation between the TailQ/TotalQ variable and
the detected number of photoelectrons of the DBLS. The band due to the neutron events can be confirmed
in the higher TailQ/TotalQ range.  The detailed analysis is ongoing;
results will be presented in the next PAC report.    

\subsection{Dynamic range of the PMT}
\indent

JSNS$^2$ is planning to use 150 10-inch photomultiplier tubes (PMTs).
A candidate PMT is R7081 of Hamamatsu Photonics, which is used in the Double Chooz and RENO
reactor $\theta_{13}$ experiments. 
 The main difference of JSNS$^2$ from Double Chooz or RENO is the light yield. 
 For the case of JSNS$^2$, the maximum neutrino energy is $53$~MeV, 10 times higher than that of
 typical reactor neutrinos. 
 The liquid scintillator is contained in an acrylic vessel and 
 the PMTs will be submerged in non-Gd loaded scintillator outside the acrylic
 vessel.
 If a neutrino signal happens in the liquid scintillator just in front of a PMT, the PMT shall
 receive scintillation light as high as $\sim$ 1,000~p.e.
 equivalent.
 On the other hand, the lowest signal energy to detect is a few MeV and if such event happens far
 from a PMT, the PMT shall receive 1 p.e. or less. 
 Since JSNS$^2$ will perform the spectrum analysis of the neutrino signal to separate the
 intrinsic $\bar{\nu}_e$ background from $\mu^-$ decay at rest, the PMT must have a
 linear dynamic range from 1~p.e. to $\sim$1,000~p.e. 
 
\begin{figure}[htbp]
  \begin{center}
    \includegraphics[keepaspectratio=true,width=140mm]
    {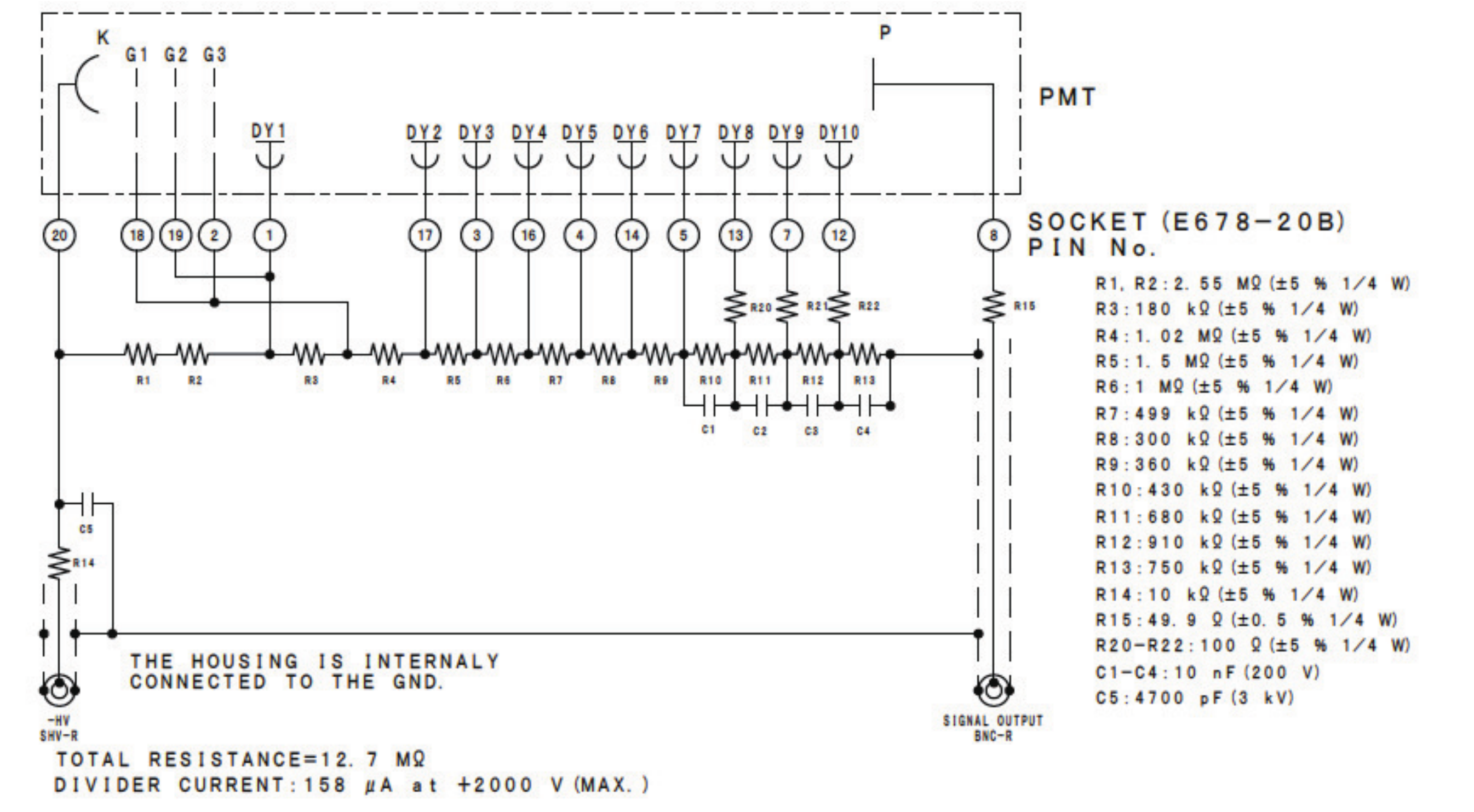}
    \caption{The diagram of the tapered circuit.  }
    \label{fig:BaseCircuit}
    \end{center}
\end{figure}

\begin{figure}[htbp]
  \begin{center}
    \includegraphics[keepaspectratio=true,width=90mm]
    {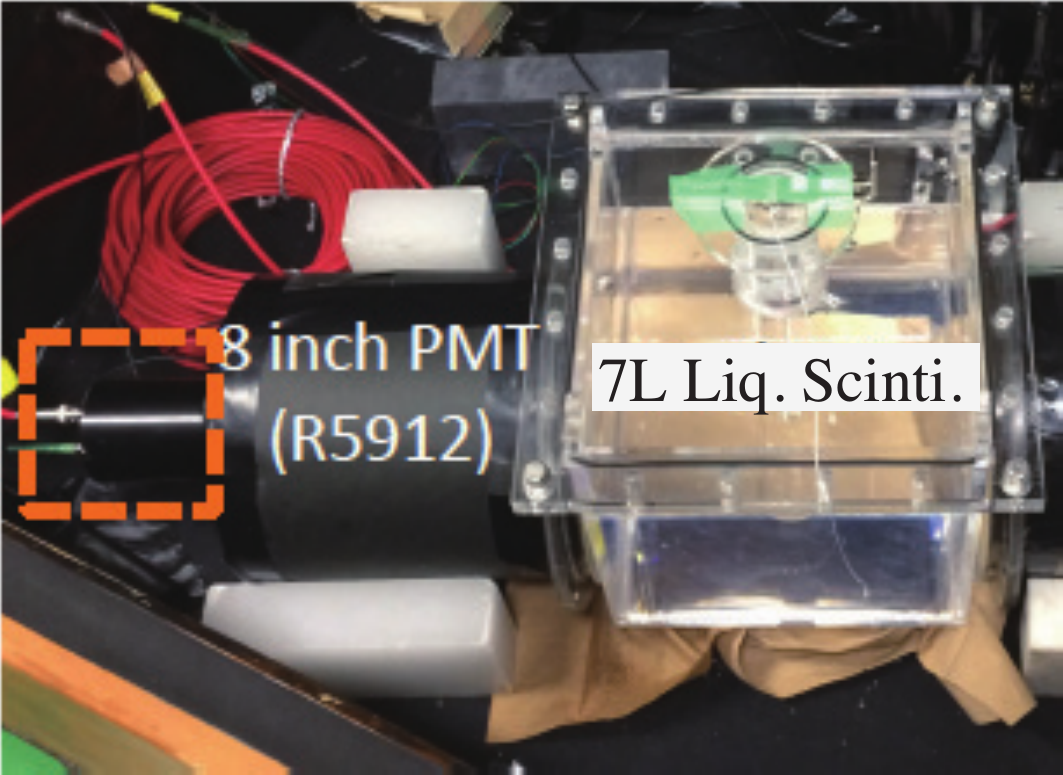}
    \caption{\setlength{\baselineskip}{4mm}
      The set up for the gain measurement.
    The dashed box shows the base socket which was swapped for regular and the tapered divider circuits.}
    \label{fig:TestEquip}
    \end{center}
\end{figure}

To obtain such large dynamic range, we will use a tapered base circuit as shown in Fig.~\ref{fig:BaseCircuit}. 
The voltage bias between dynodes is set higher for latter dynodes to reduce the space charge effect. 
The linearity of the gain was measured by using 7~liter liquid scintillator and 8~inch PMT as shown in
Fig.~\ref{fig:TestEquip}. 
 For the test, the liquid scintillator was contained in an UV-transparent acrylic vessel that was readout by an 8~inch PMT.  
 Two base circuits, one as the regular divider and the other as the tapered divider, were prepared and the
 base sockets were swapped to compare the measurement results. 
A 500~MHz, 14bit flash ADC (CAEN V1730) was used to readout the pulse shape.  

At first, one p.e. peak was measured by using a triggered LED to obtain the absolute gain. 
The gain of the PMT was set to $1.4\times 10^{6}$ for a 1~p.e. signal.
The cosmic-ray signal was measured to obtain the large amplitude signals. 
A saturation effect was measured using the ratio of the output charge (Q) and pulse height (H),
which is a good indicator of the non-linearity of the gain. 

\begin{figure}[htbp]
  \begin{center}
    \includegraphics[keepaspectratio=true,width=140mm]
    {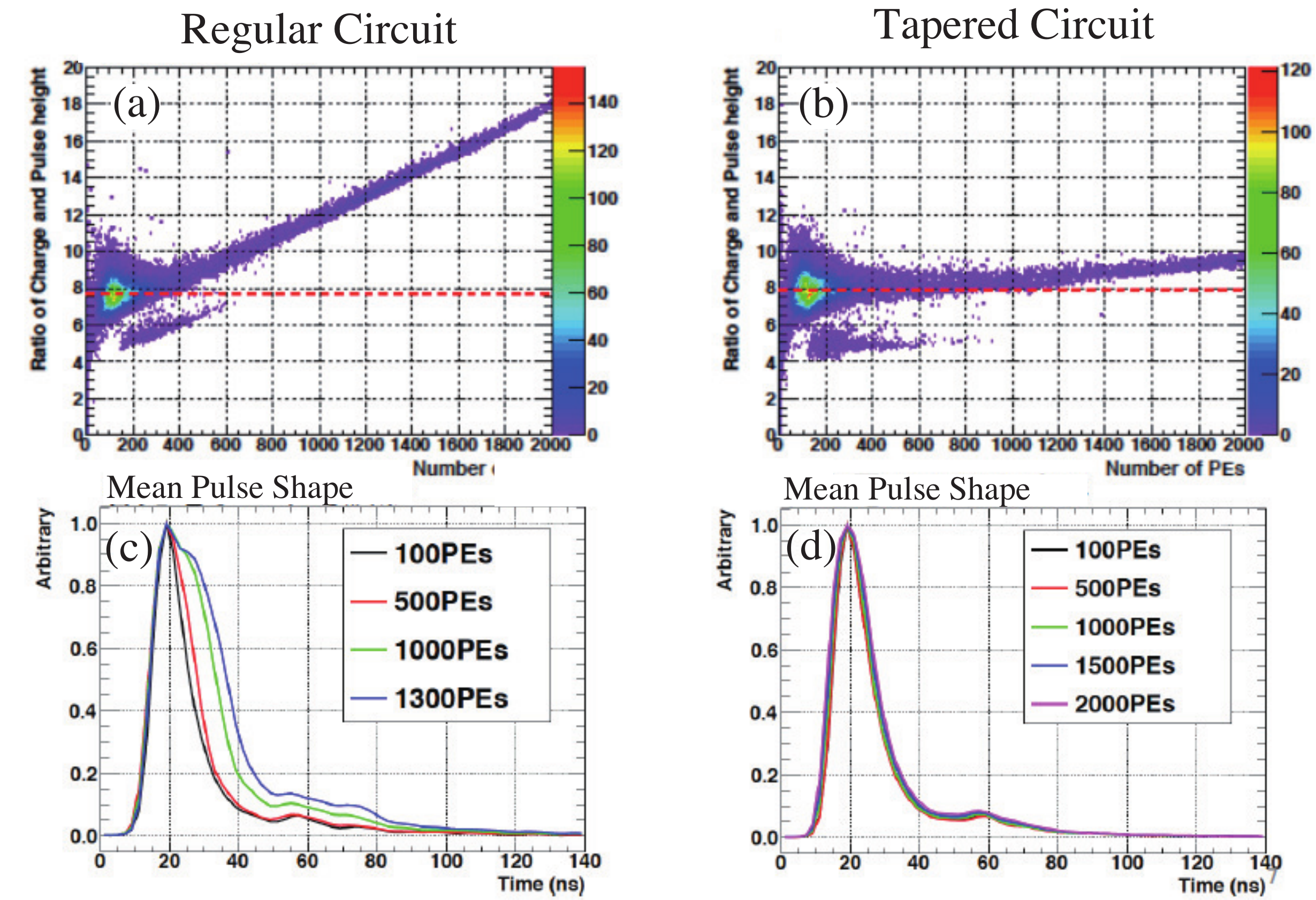}
    \caption{\setlength{\baselineskip}{4mm}
      (a)\&(c): Regular base circuit. (b)\&(d): Tapered base circuit. (a)\&(b): Q/H vs $N_{pe}$,
      (c)\&(d) Mean pulse shape.}
    \label{fig:Measurement}
    \end{center}
\end{figure}

Figures.~\ref{fig:Measurement}(a) and (b) show the Q/H ratio which indicates 
that, for the tapered circuit, the ratio is almost stable up to 1,000~p.e. while for the regular
circuit, the ratio changes from 300~p.e.

Figure~\ref{fig:Measurement}, (c) and (d), shows how the pulse shape changes depending on the input charge. 
For the PMT with the tapered base circuit, the shape does not change up to $\sim$1,000~p.e., 
which is important for the $n/\gamma$ pulse shape discrimination analysis. 

\section{Software efforts to show the fast neutron rejection}
\label{SEC:SW}
\indent

\begin{figure}[h]
\begin{centering}
\includegraphics[height=4.8in]{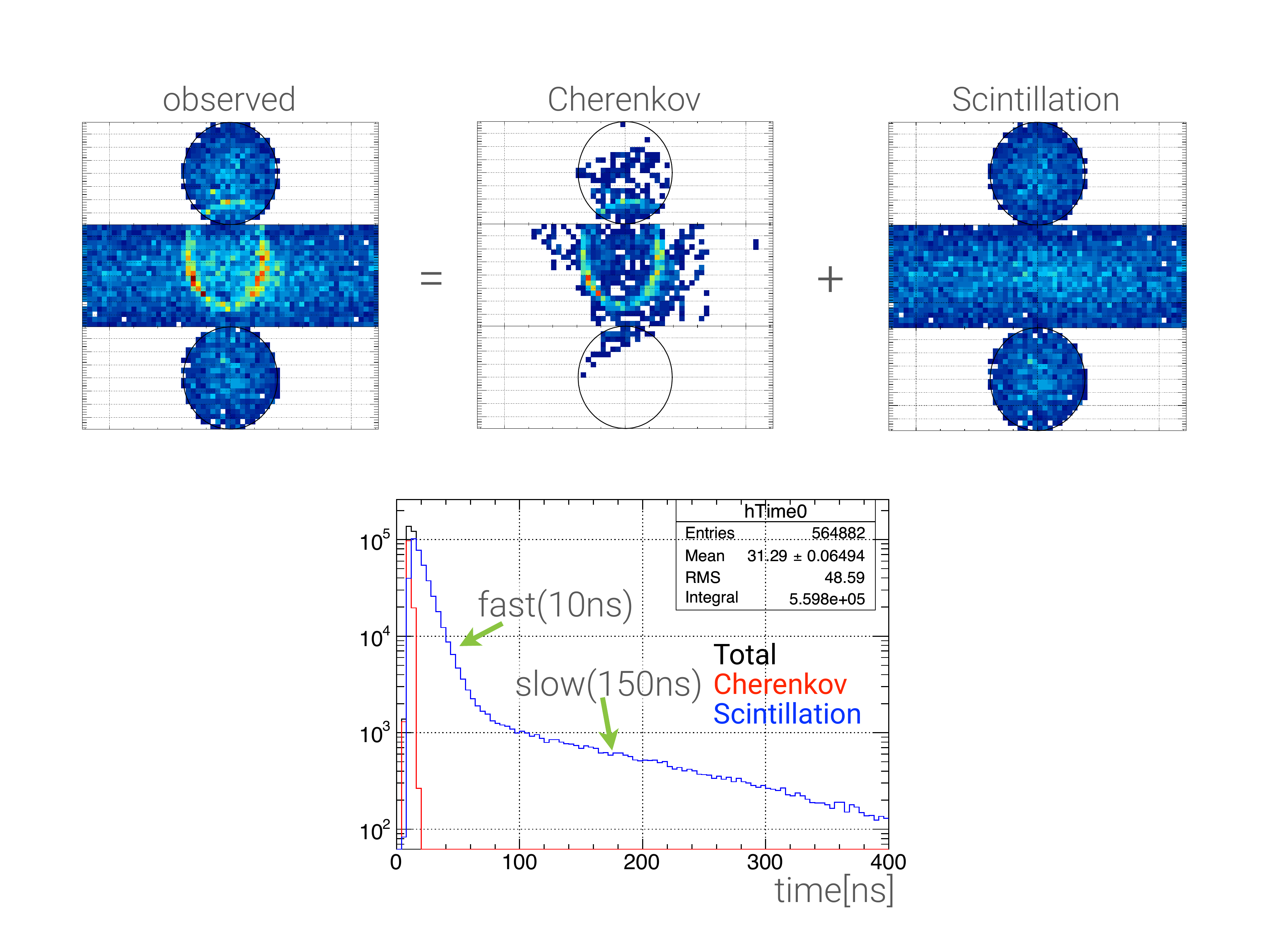} 
\vspace{-.7cm}
\caption{\setlength{\baselineskip}{4mm}
  Top: An event display featuring a simulated 50~MeV positron in the JSNS$^2$ cylindrical volume with the
  dilution factor of 10 for scintillation light. Bottom: The timing characteristics of the event.
  The Cherenkov and both the slow and fast scintillation contributions can be seen. }
  	\label{fig:evdisp}
\end{centering}
\end{figure}

At present, the most pressing issue in the JSNS$^2$ design is identifying and rejecting background fast
neutron events when selecting signal neutrinos. Cosmic-induced neutrons, rather than beam-induced neutrons,
are the primary concern. PSD and Cherenkov light can be used to differentiate the two classes of events.
However, this rejection is strongly dependent on a number of variables, including: liquid scintillator
properties, photo-coverage, light detection technology, tank geometry, veto layer(s), electronics
response, dynamic range, and timing resolution, among others. These JSNS$^2$ detector design considerations
are studied, in part, with a dedicated simulation package based on Geant4~\cite{geant4}.
An example event display is shown in Fig.~\ref{fig:evdisp} along with the timing characteristics of the
event. The figure shows a simulated 50~MeV positron in the JSNS$^2$ cylindrical volume with the dilution
factor of 10 for scintillation light. The Cherenkov ring and scintillation light components can be
easily seen. 

The simulation is tuned and validated using data taken in a number of test stands at KEK and Tohoku University.
For example, Fig.~\ref{fig_timing} shows the results of the simulation as compared to the KEK test stand
(Fig.~\ref{Fig:timing}) in consideration of Cherenkov and scintillation light. Another aspect of the
simulation involves studying PSD for separating neutron-induced proton and nuclear recoil events
from positron-like events. This discrimination is based on the shape of the waveform. Neutron-induced
event waveforms tend to have a larger tail and smaller peak amplitude than positron events. Data from the
test stand at Tohoku University (See Section~\ref{SEC:PSD} for more details), featuring small volumes of
liquid scintillator and a $^{241}$Am$^{9}$Be $\gamma$ and neutron source,  are used to develop and tune a
simulation of the interactions inside liquid scintillator as well as the PMT and electronics response
in simulating waveforms properly. Figure~\ref{FIG:PSDJSNS} shows a Monte Carlo and data comparison for
the few-MeV gamma (electron) and neutrons. Also shown is the Monte Carlo expectation for the pulse shapes
all the way up to 50~MeV, which is most relevant for JSNS$^2$. As can be seen, pulse shape discrimination
becomes more powerful for higher energy species.

The ability to differentiate neutron-induced events from positrons appears to be quite promising using both
Cherenkov-scintillation and pulse shape discrimination techniques. The software simulations continue to
be expanded in scope and tuned to existing data. The 70~MeV neutron beam test data with the Tokoku
University liquid scintillator volumes will allow the simulation to be further tuned and enhanced at
energies most representative for JSNS$^2$. 

Reconstruction software work is also ongoing concurrently with the simulation efforts. The event vertex
determination is based on a maximum likelihood algorithm using charge and time information. Both a point
source event and an extended distribution can be considered when forming the event likelihood. The
reconstructed energy is determined based on the number of photoelectrons, timing, and the vertex
determination from the likelihood algorithm above. Optimizing the reconstruction, which is highly
dependent on (e.g) non-linearities, light sensor properties, and liquid scintillator properties,
involves working closely with the simulation software as well.

\section{Summary and Plan}  
\indent

The JSNS$^2$ collaboration is making continuous progress to
produce a TDR.

For the detector location,
there were many intensive discussions between the MLF and the JSNS$^2$,
on the safety requirements for the treatment of more than 100~tons of liquid
scintillator,
the interference between areas for the irradiated materials storage and
operation of
liquid scintillator, the weight load on the MLF building due to the detectors,
leakage protection and the safety 
of the liquid scintillators, slow monitor and controls, ground sinkage of the
MLF building due to the weight of the detectors, and procedures for bringing the
detectors outside of the building. This interim report will be presented to the PAC,
and further investigation will be done to produce the TDR.

JSNS$^2$ detector R$\&$D has also made excellent progress.
For the rejection of fast neutron events 
two methods are under study. One is using Cherenkov light and the other is using PSD.
For the Cherenkov light technique, we checked the performance
of the liquid scintillator made by LAB+0.03~g/L b-PBD. This liquid scintillator
has excellent properties compared to the liquid scintillator comprising mineral oil
plus 0.03 g/L b-PBD, in consideration of the light emission time for distinguishing between
Cherenkov and scintillation light.
For the method using PSD, we employed a test-beam using beam fast neutrons at Tohoku University.
The analysis is on-going. 
The dynamic range of the 10'' PMT was improved by more than three times due to
the change of a register chain of the PMT base circuit.
This improvement allows JSNS$^2$ to have 
a good energy resolution up to $\sim$ 50 MeV.

In order to transform this R$\&$D information into the detector performance
of the JSNS$^2$, a software framework has been set up using Geant4.
The data of the Cherenkov test using the prototype is reasonably
reproduced by Geant4. The PSD performance for the neutron-proton elastic
events in the real size detector was estimated using the simulation. These are
also on-going efforts.

In summary, the continued JSNS$^2$ progress indicates the TDR will be produced on schedule,
within about one year from now. 

\section{Acknowledgements}
\indent

We warmly thank the MLF people, especially, Dr. Futakawa, J-PARC sub-director
and Prof. Kanaya, MLF Division leader, 
the neutron source group, muon group and the user facility group for their 
support.
For the discussion between the MLF and the JSNS$^2$ experiment, we appreciate
the help from Katsuo Tokushuku (KEK IPNS director), Naohito Saito
(J-PARC director),
Tetsuro Ishii (J-PARC sub-director), Takashi Kobayashi (KEK IPNS sub-director)
and the J-PARC safety Division in addition to the MLF people mentioned above
to set up constructive discussions.

Finally, we also acknowledge the support from grant-in-aid, J-PARC and KEK. 

\newpage
\appendix
\section{Additional Physics Study under Consideration}

\subsection{Physics with neutrinos from charged kaon decay-at-rest}
\indent

JSNS$^2$ has the unique ability to precisely measure monoenergetic 236~MeV neutrinos from charged kaon decay-at-rest ($K^+ \rightarrow \mu^+ \nu_\mu$; BR=63.5\%~\cite{pdg}) for the first time. These neutrinos represent (1) an unprecedented weak-interaction-only, known energy probe of the nucleus, (2) a standard candle for developing a thorough understanding of the neutrino interaction and cross sections critical for future long baseline neutrino experiments, and (3) a source for a sterile neutrino search using electron neutrino appearance~\cite{kdar1,kdar2}. These neutrinos have also been cited as important for probing muon neutrino disappearance at short baseline~\cite{kdar3} and as a possible dark matter annihilation signature~\cite{kumar}. 

Despite the importance of the kaon decay-at-rest (KDAR) neutrino across multiple aspects of particle and nuclear physics, these neutrinos have never been studied or even identified before. The decay-in-flight neutrino ``background" in conventional beamlines drowns out the KDAR signal in such experiments. Decay-at-rest sources of neutrinos, most notably spallation neutron sources, are excellent locations for studying KDAR due to their minimal decay-in-flight background and intense beams. However, the historically most intense spallation sources have been too low energy to produce kaons readily. The J-PARC MLF 3~GeV primary proton energy is sufficient to produce kaons efficiently and, also in consideration of the facility's beam intensity (eventually 1~MW, currently 500~kW~\cite{mlf_news}), represents the best facility in the world to accomplish this physics. The KDAR neutrino can easily be seen in Fig.~\ref{fig:sense}, which shows the neutrino flux at the J-PARC MLF source.  

JSNS$^2$ expects to collect a sample of between 150,000 and 300,000 $\nu_\mu$ charged current events in 50~tons of fiducial volume in its 5 year run\footnote{
  \setlength{\baselineskip}{4mm} The large variation in the expected number of events is due to the highly uncertain kaon production at this energy. The lower and upper bounds come from Geant4~\cite{geant4} and MARS~\cite{mars} predictions, respectively.}. These events ($\nu_\mu n \rightarrow \mu^-  p$ or $\nu_\mu  \mathrm{^{12}C} \rightarrow \mu^- X$) are easily identifiable due to the characteristic double coincident signal of the prompt muon plus proton(s)/nucleus followed by the muon decay electron ($\mu^- \rightarrow e^- \overline{\nu}_e \nu_\mu$) a few $\mu s$ later. 

\begin{figure}[htpb]
\centering
\begin{minipage}{.6\linewidth}
  \includegraphics[width=\linewidth]{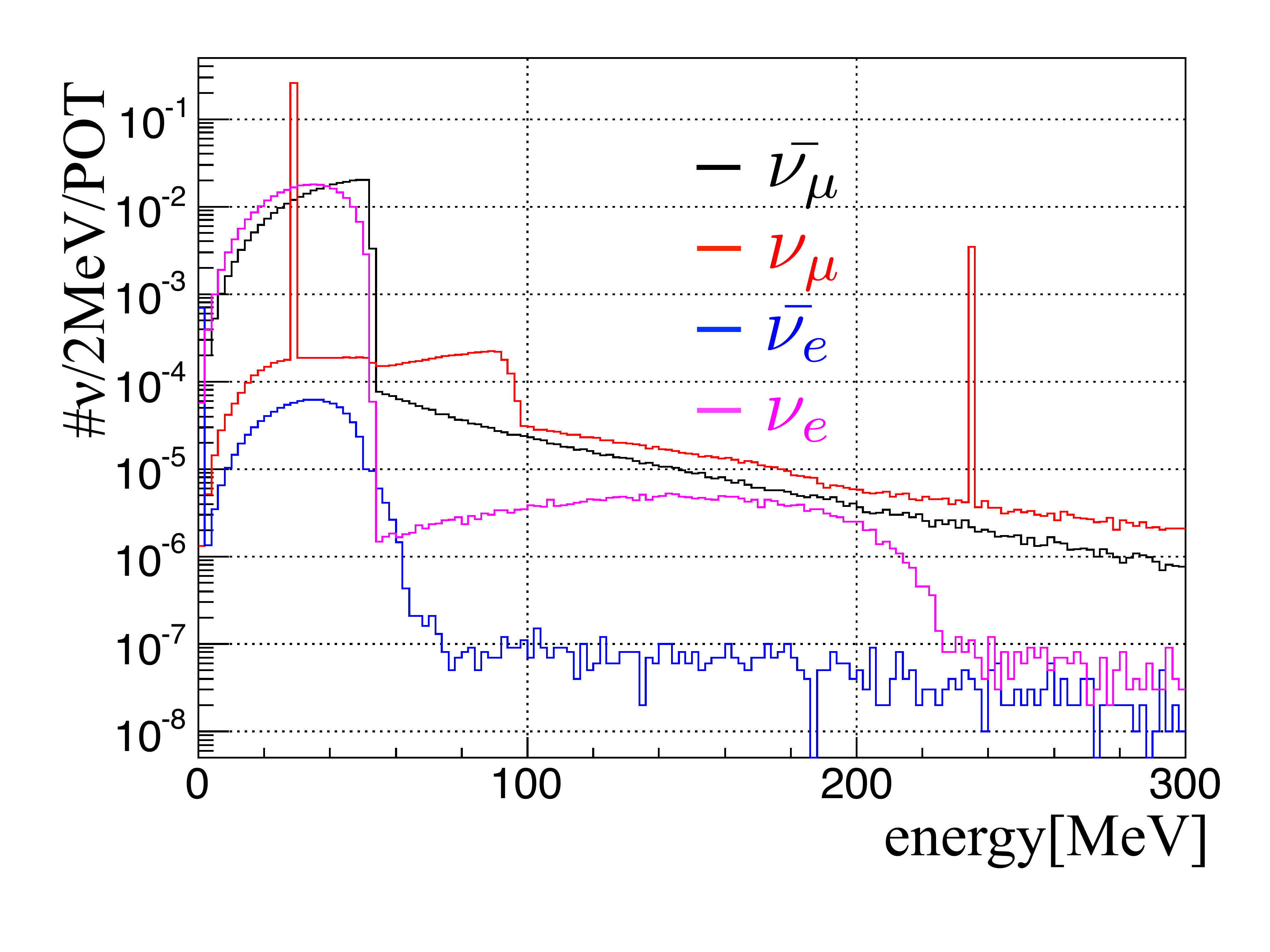}
\end{minipage}
\vspace{-.3cm}
\caption{\setlength{\baselineskip}{4mm}
  The neutrino flux at the J-PARC MLF source. The 236~MeV muon neutrino from charged kaon decay-at-rest can easily be seen.}
\label{fig:sense}
\end{figure}

\begin{figure}[htpb]
\begin{centering}
\includegraphics[height=4.8in]{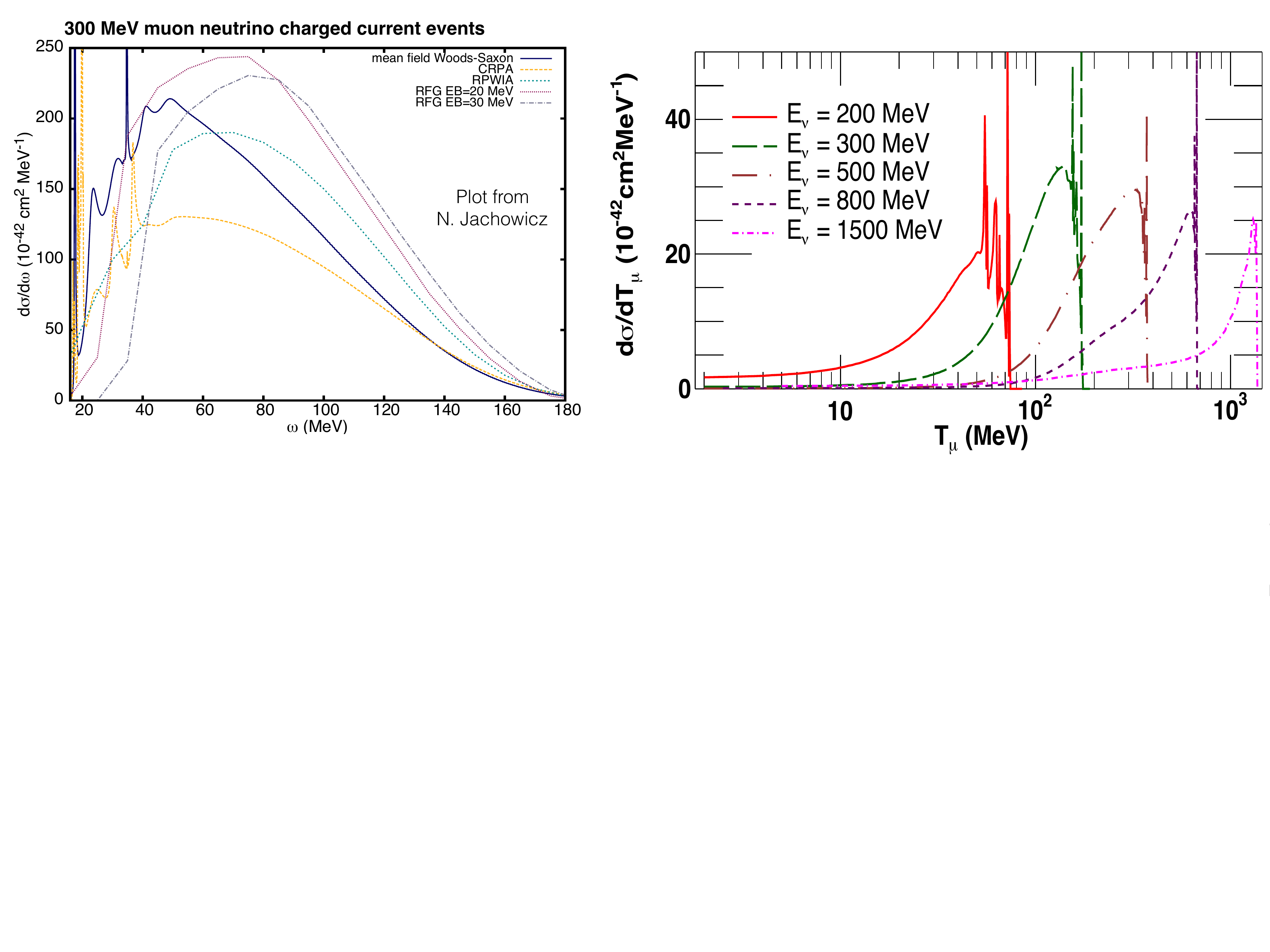} 
\vspace{-6.7cm}
\caption{\setlength{\baselineskip}{4mm}
  Left: The differential cross section in terms of energy transfer ($\omega=E_\nu-E_\mu$) for 300~MeV $\nu_\mu$ CC scattering on carbon. Predictions from various models are shown. This plot is adapted from Ref.~\cite{natalie}. Right: The differential cross section in terms of muon kinetic energy for various neutrino energies within the Continuum Random Phase Approximation model; this plot is adapted from Ref.~\cite{crpa}.}
  	\label{fig:kinematics1}
\end{centering}
\end{figure}

The known energy KDAR neutrinos provide the exclusive tool, for the first time, to study nuclear structure and the axial vector component of the interaction using electron scattering variables such as $\omega$ ($\omega=E_\nu-E_\mu$). The importance of this unique access to the nucleus is potentially far-reaching. For example, a double differential cross section measurement in terms of $\omega$ vs. $Q^2$ allows one to distinguish effects of the form factors, which depend only on $Q^2$, and of the nuclear model, which depends on both. Figure~\ref{fig:kinematics1} (left) shows a number of model predictions for the differential cross section in terms of energy transfer for 300~MeV $\nu_\mu$ CC scattering on carbon. The disagreement between the models, in terms of both shape and normalization, is striking. Notably, the JSNS$^2$ muon energy resolution may allow the nuclear resonances, easily seen in Fig.~\ref{fig:kinematics1} (right), to be measured via neutrino scattering. The KDAR neutrino is likely the only way to study these excitations with neutrino scattering and, in general, to validate/refute these models in the $<400$~MeV neutrino energy range (see, e.g., Ref.~\cite{teppei}).

Along with studying nuclear physics relevant for future neutrino experiments, the large sample of KDAR muon neutrinos collected with JSNS$^2$ will provide a standard candle for understanding the neutrino energy reconstruction and outgoing lepton kinematics in the 100s-of-MeV neutrino energy region. While the KDAR neutrino is simply not relevant for experiments featuring significantly higher neutrino energies, most notably for MINOS, NOvA and DUNE~\cite{minos,nova,dune}, it is highly relevant for experiments with a large or majority fraction of few-hundred-MeV neutrinos, for example, T2K~\cite{t2knim}, MOMENT~\cite{moment}, the European Spallation Source Neutrino Super Beam (ESS$\nu$SB)~\cite{essnu}, and a CERN-SPL-based neutrino beam CP search~\cite{beta_spl}. In particular, MOMENT and ESS$\nu$SB both feature $\nu_\mu$ spectra which peak at about 200-250~MeV.

The KDAR neutrino can also be used to search for electron neutrino appearance ($\nu_\mu \rightarrow \nu_e$) for providing a probe of the sterile neutrino that will be highly complementary to the JSNS$^2$ IBD search ($\overline{\nu}_\mu \rightarrow \overline{\nu}_e$). The advantage of the KDAR technique over other sterile neutrino searches is that the signal energy (236~MeV) is known exactly. A background measurement on either side of the signal energy window around 236~MeV can allow an interpolated determination of the expected background in the signal region with high precision. However, as compared to $\nu_\mu$ CC interactions, $\nu_e$ events ($\nu_e n \rightarrow e^-  p$ or or $\nu_e  \mathrm{^{12}C} \rightarrow e^- X$) are more challenging to identify over background, since they do not feature a double coincidence signal. While the KDAR $\nu_e$ events are expected to be distinct, in the sense that their reconstructed energy will lie close to 236~MeV, beam-induced neutrons can interact inside of the detector to produce an energetic single flash of light (e.g. a proton), mimicking a 236~MeV $\nu_e$ event. Pulse shape discrimination and particle identification via Cherenkov light can be used to mitigate this background, but the background event rate expectation remains significant. This is worrisome because the oscillated signal expectation is $<100$~events in consideration of the global best fit region at high-$\Delta m^2$. The possibility of probing $\nu_e$ appearance using KDAR neutrinos at the MLF remains an intriguing possibility, however, especially given strong Cherenkov-scintillation separation, pulse shape discrimination, and/or additional shielding.

The KDAR neutrino opens up new avenues for research in neutrino oscillation, interaction, and nuclear physics, and the importance of these measurements is clear. Perhaps most intriguing, the nucleus has simply never been studied using a known energy, weak-interaction-only probe and KDAR provides the exclusive technique to explore this frontier. JSNS$^2$ represents the world's best hope to take advantage of the KDAR neutrino in the near future. Other existing facilities worldwide simply cannot match the large expected KDAR signal and small expected decay-in-flight background rate at the J-PARC MLF.

\subsection{\setlength{\baselineskip}{6mm} Measurement of neutrino-induced 
nuclear reaction cross sections}

\subsubsection{Physics Motivations}
\indent

Stars with initial masses greater than 8$\sim$10 times the
solar mass are expected to end their lives with core collapsing
supernova explosions (Type-II SNe). When such a massive star 
exhausts its nuclear fuel at the center, it produces no more heat 
to sustain its own weight, and materials in the outer layers begin to 
fall into the central core. The density and the temperature of 
the core rapidly increases, and the core is photo-dissociated to 
a mixture of nucleons and light nuclei, which absorb free electrons 
by emitting neutrinos. Finally, a proto-neutron star is formed.
The core becomes so hard that the falling materials are repelled 
at the surface of the core. The repelled matter collides with the 
subsequently falling matter, generating an outgoing shockwave. 
It was found through a number of simulations on Type-II SNe within 
the spherically symmetric one-dimensional or two-dimensional models 
that the outgoing shockwave cannot get the kinetic energy of about 
10$^{51}$ erg which is necessary to propagate to infinity~\cite{SN1}. 
In order to solve the problem, many effects such as stellar rotation, the magnetic field and the neutrino-nucleus 
interactions have been studied~\cite{SN2}, and now the delayed 
explosion model is considered as one of the promising and realistic 
models for Type-II SNe~\cite{SN3}. 
In this model, the kinetic energy needed for the explosion is 
considered to be supplied by the nuclear reactions between the 
matter in the shockwave and the neutrinos which carry almost 
99$\%$ of the initial gravitational energy of the star, which is 
of the order of 10$^{53}$ erg. From precise simulations of the 
dynamics of Type-II SNe, it was found that the neutrinos are 
trapped in the core, and are released after about a hundred 
milliseconds. Then they overtake the outgoing shockwave, 
supplying an additional kinetic energy via the neutrino-nucleus 
interactions. Since the region where the shockwave stalls is 
composed of light nuclei~\cite{SN4}, the neutrino-induced 
nuclear reactions on light nuclei should play critical roles 
in the dynamics of explosions. From the one-dimensional and
two-dimensional simulations for the explosion energy in Type-II 
SNe for different neutrino luminosities and fixed neutrino-nucleus 
reaction rates, it was found that the enhancement of the neutrino 
luminosity by 10$\sim$15$\%$ leads to an increase of the order 
of 10$^{50}\sim$10$^{51}$ erg/s in the explosion energy~\cite{SN5}. 
It suggests the enhancement of the neutrino-nucleus reaction 
rates with fixed neutrino luminosity also should give the same 
magnitude of increase in the explosion energy. Therefore the 
neutrino-nucleus reaction rates should be known with accuracy 
better than about 10$\%$ to evaluate the effect of neutrinos 
to the explosion. 

Another important role of the neutrino-induced nuclear reactions 
concerns the r-process nucleosynthesis. A recent scenario of the 
r-process nucleosynthesis assumes the formation of the 
high-entropy neutron-rich gas, called the neutrino-driven wind, 
in the atmosphere of a nascent neutron star by the neutrino-induced 
spallation reactions, and the synthesis of heavy elements from protons 
and neutrons up to the nuclides with the mass number of 200$\sim$250 
within about one second~\cite{SN6,SN7,SN8}. 
This scenario is preferred, because it does not 
require the existence of the seed nuclei like iron, and naturally 
explains the universality of the r-elements, i.e. the similarity 
found in the r-element abundances in stars with different 
metallicities. More recently, it was pointed out that the 
neutrino-induced spallation reactions on light nuclei may 
efficiently produce lithium and boron in the oxygen/carbon layer of 
a Type-II supernova~\cite{SN9}. Such a process is interesting as 
a possible source of Li-Be-B in addition to the spallation 
by cosmic rays, and also as a new method to constrain the parameter 
of the flavor oscillation of the neutrinos. 
For more precise calculations, it is necessary to use accurate 
rates of the neutrino-induced nuclear reactions. 

\subsubsection{\setlength{\baselineskip}{6mm} Current status of 
data for neutrino-induced nuclear reaction cross sections}

\begin{figure}[htpb]
 \centering
 \includegraphics[width=0.6 \textwidth]{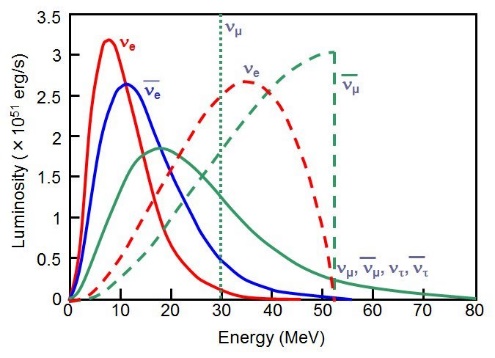}
\caption{\setlength{\baselineskip}{4mm}
Energy spectra of neutrinos from Type-II SNe (solid curves) 
and DAR pions and muons (dashed curves, arbitrary units in 
luminosity). 
}
 \label{Fig:SN_DAR}
\end{figure}

So far, experimental data of the neutrino-induced nuclear reaction 
cross sections have been obtained by using neutrinos produced by 
accelerators or radioactive isotopes. Decay-at-rest (DAR) 
neutrinos from stopped pions and muons generated with high-energy 
accelerators are very useful for studies of the nuclear reactions 
induced by supernova (SN) neutrinos, because their energy spectra 
overlap with those of SN neutrinos as shown by 
Fig.~\ref{Fig:SN_DAR}. 

Table~\ref{TAB:SN1} shows the list of the presently available 
experimental data for neutrino-induced nuclear reaction 
cross sections. 

\begin{table}[htbb]
  \caption{\setlength{\baselineskip}{4mm}
    Summary of the existing data of the neutrino-induced 
    nuclear reaction cross sections.}
\begin{center}
\begin{tabular}{|l|c|c|c|} \hline
Reaction & Neutrino Source & Accuracy & Reference \\\hline
$^{12}$C($\nu_e$,e$^-)^{12}$N$_{g.s.}$ & Accelerator $\nu$ & $\sim$10$\%$ & \cite{SN10}\cite{SN11} \\\hline
$^{12}$C($\nu_e$,e$^-)^{12}$N$^{*}$   & Accelerator $\nu$ & $\sim$15$\%$ & \cite{SN10}\cite{SN11} \\\hline
$^{12}$C($\nu$,$\nu^{'})^{12}$C$(1^+1)$ & Accelerator $\nu$ & $\sim$20$\%$ & \cite{SN10}\cite{SN11} \\\hline
$^{13}$C($\nu_e$,e$^-)^{13}$N & Accelerator $\nu$ & 76$\%$ & \cite{SN10} \\\hline
$^{56}$Fe($\nu_e$,e$^-)^{56}$Ni & Accelerator $\nu$ & 37$\%$ & \cite{SN10} \\\hline
$^{71}$Ga($\nu_e$,e$^-)^{71}$Ge & RI ($^{51}$Cr)     & 11$\%$ & \cite{SN12}\cite{SN13} \\\hline
$^{127}$I($\nu_e$,e$^-)^{127}$Xe & Accelerator $\nu$ & 33$\%$ & \cite{SN14} \\\hline
\end{tabular}
\end{center}
\label{TAB:SN1}
\end{table}

As shown in Table~\ref{TAB:SN1}, even in the case of the best studied 
$^{12}$C, the accuracy of the experimental data is not better than 
the one required in the SN simulations, and therefore new experimental 
data with better accuracy are still needed. JSNS$^2$ is expected to 
provide an experimental opportunity to measure the neutrino-induced 
nuclear reaction cross sections with better accuracy due to a 
high-intensity neutrino beam from the J-PARC/MLF and a 
high-sensitivity/high-resolution detector system. 

\subsubsection{\setlength{\baselineskip}{6mm} Plan for measurement of 
the neutrino-induced nuclear reaction cross sections in JSNS$^2$ project}

\subsubsection*{\setlength{\baselineskip}{6mm} Background rejection 
and statistical accuracy}
\indent

Assuming the neutrino flux of J-PARC/MLF to be 
$\sim$1.6$\times$10$^{14}$ /cm$^2$/y/MW, one needs the number 
of the target nucleus of at least $\sim$10$^{29}$ atoms or the target 
quantity of more than a few tons to measure the cross section in the 
range of about 10$^{-41}$ cm$^2$ with a statistical error smaller 
than 10$\%$. Therefore $^{12}$C contained in the LAB is a 
candidate for the target in the present design of the JSNS$^2$ detector. 
  Since the energies of DAR neutrinos are below 52.8~MeV as shown in 
Fig.~\ref{Fig:SN_DAR}, possible reaction channels are 
the $^{12}$C($\nu_e$,e$^-)^{12}$N reaction caused by the charged 
current (CC) and the neutrino inelastic scattering via the neutral 
current (NC). The JSNS$^2$ detector is not sensitive to low energy 
protons from the $^{12}$C($\nu$,$\nu^{’}$p)$^{11}$B reaction, and 
only the $^{12}$C($\nu_e$,e$^-$)$^{12}$N and 
$^{12}$C($\nu$,$\nu^{’})^{12}$C$^*$ reactions will be measured 
through detection of emitted electrons or $\gamma$-rays. According 
to the previous work in these areas, the background is supposed to be dominated 
by the following types; \\
Type-A:  high energy muons and their decay particles 
(Michel electrons), \\
Type-B:  $\beta$-decay of $^{12}$B following the 
$^{12}$C($\mu^{-}$,$\nu_{\mu})^{12}$B reaction, and \\
Type-C:  high energy $\gamma$-rays and neutrons induced by 
cosmic rays. \\

\begin{figure}[htpb]
 \centering
 \includegraphics[width=0.6 \textwidth]{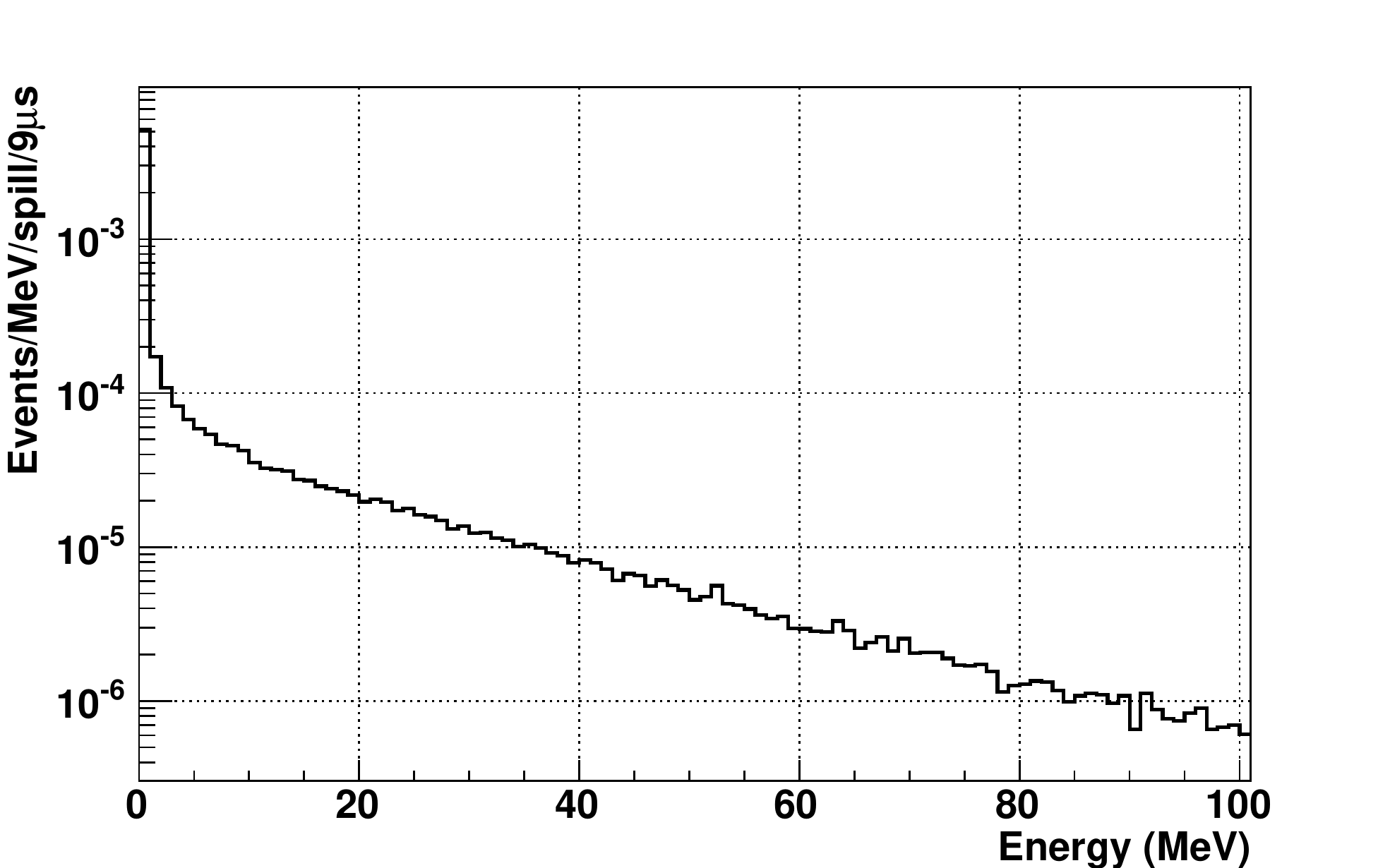}
\caption{\setlength{\baselineskip}{4mm}
The energy spectrum of the expected background of the JSNS$^2$ detector 
with the charged-veto cut and the Michel-electron cut (Fig.12 in
Ref~\cite{CITE:SR_14NOV}).
}
 \label{Fig:SN2}
\end{figure}

To reduce Type-A, it is effective to make an anti-coincidence by 
detecting parent muons in a time window of 20$\mu$s prior to the 
event trigger time (“charged-veto cut”) or decay electrons until 
after 9$\mu$s (“Michel electron cut”). For Type-B, the background will be 
efficiently suppressed by rejecting events with an energy 
smaller than $\sim$15 MeV, because the end point energy of the 
$^{12}$B $\beta$-decay is 13.4~MeV. The counting rate of the 
residual background after those cuts can be estimated by the 
Monte Carlo simulation based on the result of the study for 
the background with use of the 500~kg plastic 
scintillator~\cite{CITE:SR_14NOV}. Figure~\ref{Fig:SN2} shows 
the expected background spectrum after the charged-veto cut and 
the Michel-electron cut. Since the end-point energy of 
$\nu_e$ and the threshold energy of the 
$^{12}$C($\nu_e$,e$^-)^{12}$N reactions are 52.8~MeV and 17.3~MeV, 
respectively, the energy window for 
$^{12}$C($\nu_e$,e$^-)^{12}$N is to be set between 20 MeV and 
40 MeV, taking into account the energy resolution of the 
JSNS$^2$ detector. The detection efficiencies with the energy 
window and the time window are determined with the Monte Carlo 
calculation as 67.4$\%$ and 74$\%$, respectively, and 
the expected counting rate of the $^{12}$C($\nu_e$,e$^-)^{12}$N 
reaction is 1513 events/5000h/MW/50t. On the other hand, the 
background rate in this energy window is estimated to be 
2.43$\times$10$^5$ events/5000h/MW, being two orders of 
magnitude larger than the expected counting rate of the 
$^{12}$C($\nu_e$,e$^-)^{12}$N reaction.

Further reduction of the background will be achieved with the 
delayed coincidence of the $^{12}$N $\beta^{+}$-decay. 
Considering the Q-value (17.4~MeV) of the $^{12}$N $\beta^{+}$-decay 
and the $\gamma-$ray background rate below $\sim$8 MeV, we will 
set the energy window from 10~MeV to 18~MeV for the delayed signal. 
To avoid the beam-related background, the beginning of the time 
window will be set at 0.1~ms from the rising edge of the proton 
beam pulse. On the other hand, the end of the time window will 
be set at 47.5~ms, which corresponds to three as long as the 
half-life (15.83~ms) of $^{12}$N, in order to acquire most of the 
positrons from $^{12}$N. 
The expected counting rates of $R_{12C}$ and $R_{BG}$ for the 
$^{12}$C($\nu_e$,e$^-)^{12}$N signal and the background due to 
the accidental coincidence, respectively, are given by the 
following equations; 

\begin{equation}
R_{12C} = N_{12C} \times \Phi_{\nu(e)} \times \sigma_{12C} \times 
\varepsilon_{p} \times \varepsilon_{d}, \nonumber
\end{equation}
\begin{equation}
R_{BG} =  R_{BG(p)} \times T_{coin} \times R_{BG(d)}, 
\end{equation}  

where $N_{12C}$, $\Phi_{\nu(e)}$ and $\sigma_{12C}$ are 
the number of target $^{12}$C nuclei, the flux of $\nu_{e}$, and 
the cross section for the $^{12}$C($\nu_e$,e$^-)^{12}$N reaction, 
respectively. $R_{BG(p)}$ and $R_{BG(d)}$ are the background 
rates for the prompt and the delayed signals, respectively. 
$T_{coin}$ denotes the width of the time window for the coincidence 
between the prompt and the delayed signals. As mentioned above, 
the background rate in the prompt signal window is estimated as 
$R_{BG(p)} =$ 2.43$\times$10$^5$ events/5000h/MW/50t. 

From our previous study, the backgrounds in those windows are 
considered to be dominated by the cosmic-ray induced $\gamma-$rays 
and neutrons. Therefore $R_{BG(d)}$ can be estimated by integrating 
the energy spectrum shown by Fig.~\ref{Fig:SN2} in the energy 
region for the delayed signal. The probability of the accidental 
coincidence is calculated as 2.3 /47.5ms/MW/50t from the 
estimated $R_{BG(d)}$, suggesting still further reduction of 
the background is necessary. Since the background events are not 
correlated with the prompt signals from the 
$^{12}$C($\nu_e$,e$^-)^{12}$N reaction, they can be reduced by 
careful analysises on spatial and time correlations between 
the prompt and the delayed signals. 

\begin{figure}[htpb]
 \centering
 \includegraphics[width=0.6 \textwidth]{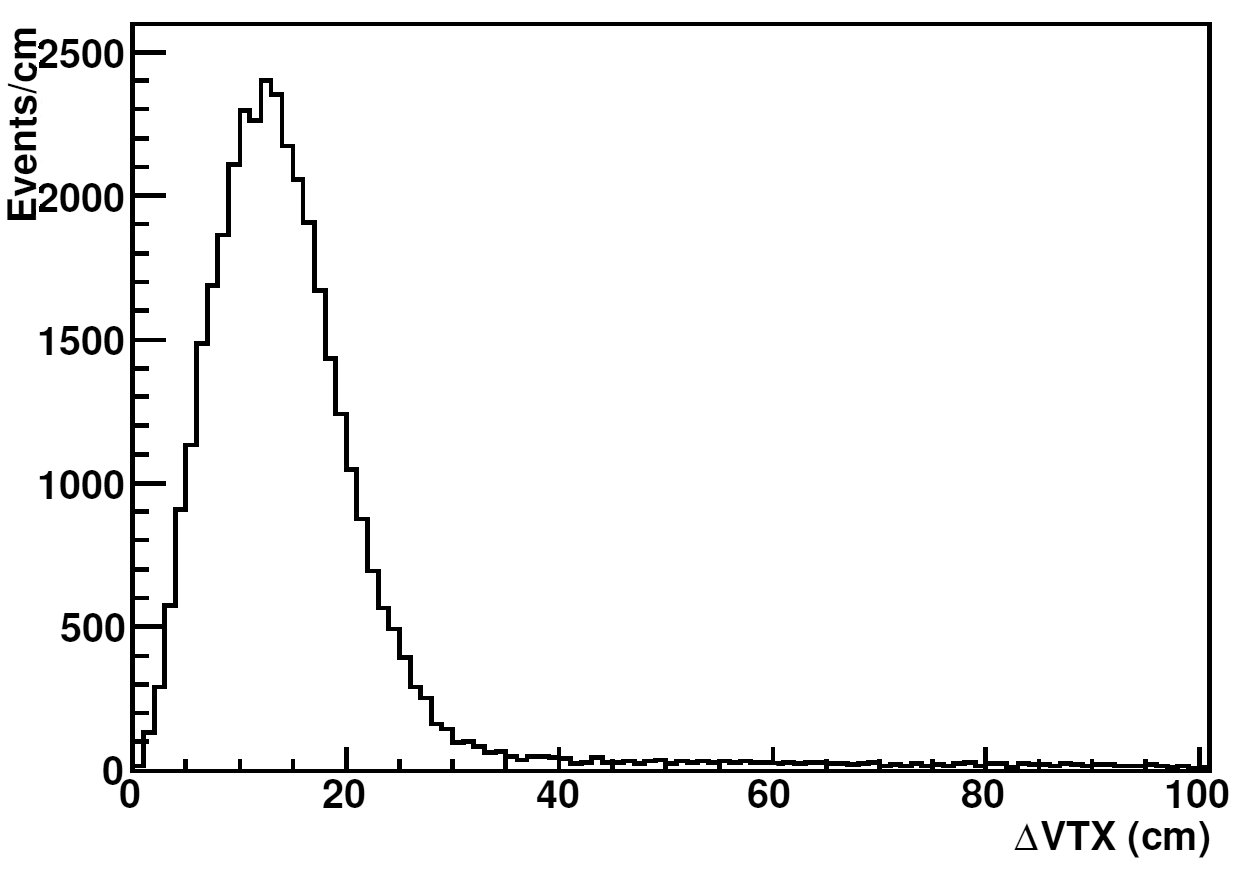}
\caption{\setlength{\baselineskip}{4mm}
Distribution of the distance ($\Delta VTX$) between prompt electron 
from $^{12}$C($\nu_e$,e$^-)^{12}$N and delayed positron 
from $^{12}$N $\beta^{+}$-decay.
}
 \label{Fig:DVTX_CNGS}
\end{figure}

Figure~\ref{Fig:DVTX_CNGS} shows the distribution of $\Delta VTX$, 
which is defined as the distance between the mean positions of 
the prompt electron from $^{12}$C($\nu_e$,e$^-)^{12}$N and the 
delayed positron from $^{12}$N $\beta^{+}$-decay estimated from 
the distributions of the photons detected with PMTs.
Setting the criterion $\Delta VTX <$ 20~cm, 82$\%$ of the 
$^{12}$C($\nu_e$,e$^-)^{12}$N events are picked up, while the 
background rate can be suppressed with a factor of $\sim$ 0.0011, 
which is basically determined by the ratio of the whole volume 
of the detector to the volume of a sphere with a radius of 20~cm. 
Concerning the time correlation, the event rates of the prompt 
electron and the delayed positrons are governed by the decays 
with half-lives of 3.17~$\mu$s and 15.83~ms, respectively, with 
respective to the primary beam pulse. The 65$\%$ fraction of the random 
background can be eliminated by the likelihood analysis on 
the time distributions of the prompt and delayed signals 
(``lifetime cut": the similar cut as mentioned in Ref~\cite{CITE:SR_14NOV}).
On the other hand, according to our previous 
simulation, 91$\%$ of the $^{12}$C($\nu_e$,e$^-)^{12}$N events 
will be kept. \\

The efficiencies and the rejection factors of the above cuts 
are summarized in Table~\ref{TAB:SN2}.

\begin{table}[htb]
  \caption{\setlength{\baselineskip}{4mm}
    Detection efficiencies and rejection factors of the cuts 
    after the charged-veto cut and the Michel-electron cut 
    evaluated with a Monte Carlo calculation. Note the rejection 
    factors with the energy windows for the prompt and delayed 
    signals have already been applied to estimate the background 
    rates $R_{BG(p)}$ and $R_{BG(d)}$ (see the text for detail).}
\begin{center}
\begin{tabular}{|c|c|c|} \hline
Cut & Efficiency & Rejection factor\\\hline \hline
Energy window (prompt 20-40MeV) & 0.674 & --- \\\hline
Time window (prompt 1-10$\mu$s) & 0.74 & --- \\\hline
Energy window (delayed 10-18MeV) & 0.42 & --- \\\hline
Time window (delayed 0.1-47.5ms) & 0.944 & --- \\\hline
$\Delta$VTX cut & 0.82 & 0.0011 \\\hline
Lifetime cut & 0.91 & 0.45 \\\hline \hline
total & 0.148 & 4.95$\times$10$^{-4}$ \\ \hline
\end{tabular}
\end{center}
\label{TAB:SN2}
\end{table}

Here it should be noted that the contribution of 
$^{12}$C($\nu_e$,e$^-)^{12}$N$^*$ can be eliminated by requiring the 
delayed coincidence of the $^{12}$N $\beta^{+}$-decay, because all 
the known excited states of $^{12}$N are located above the threshold of 
the proton emission.

Assuming the flux of $\nu_{e}$ to be 1.6$\times$10$^{14}$ /cm$^2$/y/MW 
and the $^{12}$C($\nu_e$,e$^-)^{12}$N$_{g.s.}$ cross section to be 
8.9$\times$10$^{-42}$ cm$^{2}$, the expected numbers of the events are 
estimated with the total efficiency and the total rejection factor in 
Table~\ref{TAB:SN2} as 448 /y and 138 /y for the 
$^{12}$C($\nu_e$,e$^-)^{12}$N$_{g.s.}$ reaction and the background, 
respectively, with the total detector mass of 50~t and the operation 
of J-PARC/MLF at 1~MW for 5000~h/y, leading to a very good statistical 
accuracy of 6.0$\%$ (2.7$\%$) with a one- (five-) year measurement. 

\subsubsection*{Systematic errors}
\indent

In this measurement, major sources of the systematic error are supposed 
to be the uncertainties associating the absolute flux of incident neutrinos 
and the detector response. 
In the previous experiments by the KARMEN~\cite{SN15} and the 
LSND~\cite{SN11} collaborations, the neutrino fluxes were estimated by means 
of Monte Carlo simulations using the experimentally determined ratios 
of $\mu^+$ or $\pi^+$ to the incident protons, and the uncertainties in 
the neutrino fluxes were not smaller than 7$\%$, being greater than 
statistical errors of those experiments. A more accurate determination of 
the neutrino flux may be achieved by directly measuring the rates of the 
d($\nu_e$,e$^-$p)p and d($\nu_e$,$\nu_e^{’}$n)p reactions, whose cross 
sections are known with accuracy of ~2$\%$~\cite{SN16}. This kind of 
measurement will need an additional deuterated liquid scintillation counter 
located near the spallation target of J-PARC/MLF. 
For precise understanding of the detector response, we will develop movable 
$\beta/\gamma$ sources in the active volume for calibrations of the position 
and energy measurements.

\subsubsection*{\setlength{\baselineskip}{6mm} Determination of 
$^{12}$C($\nu_e$,e$^-)^{12}$N$^*$ cross sections}

To determine the cross section for $^{12}$C($\nu_e$,e$^-)^{12}$N$^*$, 
the spectrum obtained for null delayed $\beta^+$ signal should be 
deconvoluted with the response functions for possible reaction channels 
such as $^{12}$C($\nu_e$,e$^-)^{12}$N$^*$, $^{13}$C($\nu_e$,e$^-$)X 
and $^{12}$C($\nu_e$,e$^-)^{12}$N$_{g.s}$ with missing delayed signal. 
Due to more statistics than previous experiments and precise 
calibration of the detector response, better accuracy in the analysis 
of the angular distribution is expected.

\subsubsection*{\setlength{\baselineskip}{6mm} Possibility for cross 
section measurements on neutrino-induced reactions other than 
$^{12}$C($\nu_e$,e$^-$)$^{12}$N}
\indent

As mentioned above, an accurate measurement of a neutrino-induced reaction 
cross section requires a target mass of around one ton or more and a very 
low background rate. From the viewpoint of the target mass, 
$^{12}$C($\nu,\nu^{’})^{12}$C$^*$ and $^{13}$C($\nu_e$,e$^-$)$^{13}$N are 
the candidates, but unfortunately, there is no useful decay in both 
reactions for the delayed coincidence, and therefore thick passive shields 
should be added to suppress the background due to cosmic-ray induced 
$\gamma$-rays and neutrons. Another candidate will be the 
$^{28}$Si($\nu_e$,e$^-$)$^{28}$P reaction followed by the $\beta^+$-decay 
of $^{28}$P with a Q-value of 14.35~MeV and a half-life of 270.3~ms. 
In this case the material for the liquid scintillator has to be replaced 
with a silicon compound such as silicon oil, and it should be regarded as 
one of the future extensions of JSNS$^2$ after completion of the sterile 
neutrino search.


\end{document}